\documentclass[prl,aps,twocolumn,showpacs,floatfix]{revtex4-1}
\usepackage{amsfonts}
\usepackage{amssymb}
\usepackage{hyperref}
\usepackage{graphicx}
\usepackage{dcolumn}
\usepackage{bm,amsmath,verbatim}
\usepackage{mathrsfs}
\usepackage{color}

\usepackage{amsmath,amssymb,amsthm}

\usepackage[T1]{fontenc}
\usepackage[latin9]{inputenc}
\usepackage{booktabs}

\hypersetup{colorlinks,
linkcolor=blue,          citecolor=blue,        filecolor=blue,      urlcolor=blue           }

\def\sgn{{\text{sgn\,}}}
\def\be{\begin{equation}}
\def\ee{\end{equation}}
\def\bea{\begin{eqnarray}}
\def\eea{\end{eqnarray}}
\def\bse{\begin{subequations}}
\def\ese{\end{subequations}}

\def\sgn{{\text{sgn\,}}}

\def\be{\begin{eqnarray}}
\def\ee{\end{eqnarray}}

\begin{document}

\title{Structural Disorder Induced Second-order Topological Insulators in Three Dimensions}
\author{Jiong-Hao Wang$^{1}$}
\thanks{These authors contribute equally to this work.}
\author{Yan-Bin Yang$^{1}$}
\thanks{These authors contribute equally to this work.}
\author{Ning Dai$^{1}$}
\author{Yong Xu$^{1,2}$}
\email{yongxuphy@tsinghua.edu.cn}
\affiliation{$^{1}$Center for Quantum Information, IIIS, Tsinghua University, Beijing 100084, People's Republic of China}
\affiliation{$^{2}$Shanghai Qi Zhi Institute, Shanghai 200030, People's Republic of China}

\begin{abstract}
Higher-order topological insulators are established as topological crystalline insulators protected by crystalline symmetries.
One celebrated example is the second-order topological insulator in three dimensions that hosts chiral hinge modes protected by crystalline symmetries.
Since amorphous solids are ubiquitous, it is important to ask whether such
a second-order topological insulator can exist in an amorphous system without any spatial order.
Here we predict the existence of a second-order topological
insulating phase in an amorphous system without any crystalline symmetry. Such a topological phase manifests in the winding number of the quadrupole moment, the quantized longitudinal conductance and the hinge states. Furthermore, in stark contrast to the viewpoint that structural disorder should be detrimental to the higher-order topological phase, we remarkably find that structural disorder can induce a second-order topological insulator from a topologically trivial phase
in a regular geometry. We finally demonstrate the existence of a second-order topological phase in amorphous systems with time-reversal symmetry.
\end{abstract}
\maketitle

Amorphous solid phases are ubiquitous in condensed matter systems~\cite{zallenbook}.
Since atoms in amorphous materials are randomly distributed in space, the solids
do not respect translational symmetries.
While the physics on topological phases of matter is mainly established in crystalline solids with spatial order,
it was surprisingly found that topological states can also occur in amorphous solids~\cite{Agarwala2017PRL,Chong2017PRB,Mitchell2018NP,Ojanen2018NC,Prodan2018,Yang2019PRL,Zhang2019PRB,Chern2019EPL,
Fazzio2019Nano,Corbae2019exp,Agarwala2020PRB,Ojanen2020PRR,Grushin2020arXiv,Huang2020Research,Ojanen2020arXiv,
Chong2020Light,Grushin2020review,Corbae2020arXiv,Fazzio2020arXiv}.
Recently, topological phases have been generalized to the higher-order case where
$(n-m)$-dimensional (with $1<m\le n$) gapless boundary states happen in an $n$-dimensional system~\cite{Taylor2017Science,Fritz2012PRL,ZhangFan2013PRL,Slager2015PRB,Brouwer2017PRL,FangChen2017PRL,Bernevig2018SciAdv,
 Brouwer2019PRX,Roy2019PRB,Hatsugai2019PRB,SYang2019PRL,Agarwala2020PRR,Klinovaja2020PRR,DongHui2019PRL,Yanbin2020PRR,Qibo2020PRB,
 Yang2020PRL,XRWang2020,Tao2020NJP,Tiwari2020PRL,Klinovaja2020Arxiv}.
For instance, a quadrupole topological insulator in two dimensions (2D)
can support zero-energy corner modes~\cite{Taylor2017Science,Yanbin2020PRR}. In fact, zero-energy corner modes have also been found
in a two or three dimensional amorphous lattice~\cite{Agarwala2020PRR}, which are later understood as protected by chiral symmetry~\cite{Yanbin2020Arix,Shen2020PRL}.
However, chiral symmetry is usually absent in amorphous materials.
In three dimensions (3D), a second-order topological insulator (SOTI) holding chiral hinge modes can exist
protected by the combination of the time-reversal symmetry and the four-fold rotational symmetry~\cite{Bernevig2018SciAdv}.
The requirement of the crystalline symmetry to protect a SOTI may suggest
the absence of the topological phase in amorphous systems,
despite some evidence of their robustness to weak on-site disorder~\cite{Geier2018PRB}.
Furthermore, while a previous study shows that structural disorder, disorder arising from randomly distributed atoms in space,
is detrimental to the higher-order topology protected by chiral symmetry~\cite{Agarwala2020PRR}, it is unclear whether this is the case
for the 3D second-order topology.

\begin{figure}[t]
\includegraphics[width=3.0in]{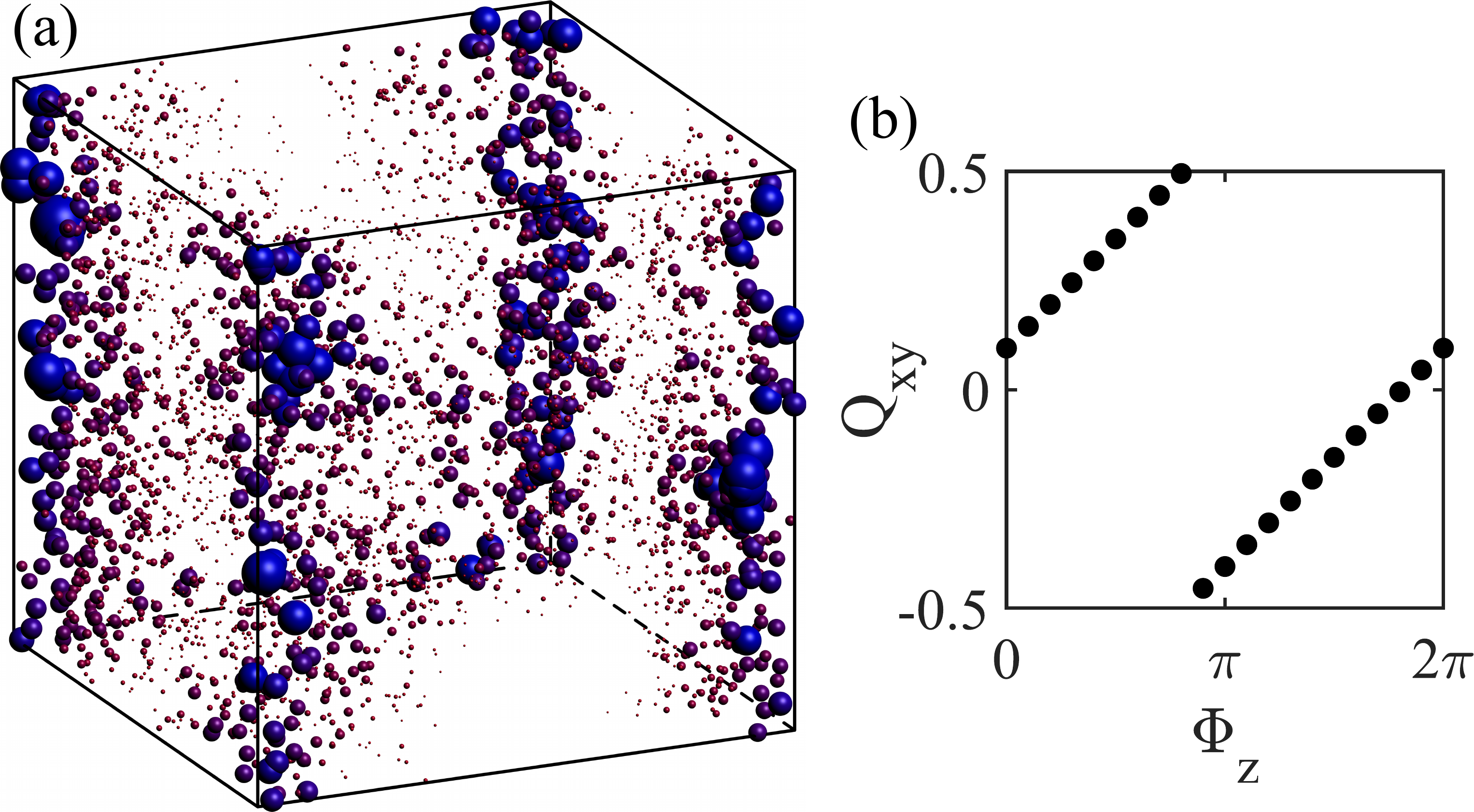}
\caption{(Color online)
(a) In-gap hinge states with the size of each sphere indicating the sum of local densities of four eigenstates closest to zero energy at that site.
(b) The quadrupole moment $Q_{xy}$ with respect to an inserted flux $\Phi_z$, reflecting that the winding number $W_Q=1$ defined in (\ref{winding1}).
Here, we consider a typical sample of a SOTI on an amorphous lattice of size $L=20$ for $M=-3$ in Hamiltonian (\ref{Ham1}).
}
\label{Fig1}
\end{figure}

In this work, we theoretically demonstrate the existence of a second-order topological insulating phase in a 3D
random lattice model without any symmetry.
We find that despite the complete breaking of the translational symmetry,
the amorphous system can still exhibit nonzero quantized winding number of the quadrupole moment (see Fig.~\ref{Fig1}) associated
with nonzero quantized longitudinal conductances $2e^2/h$. The two-terminal conductance is contributed by the chiral
modes localized at the hinges (see Fig.~\ref{Fig1}) evidenced by the local density of states (LDOS).
In the 2D (or 3D) amorphous system with chiral symmetry, it has been shown that when structural disorder percolates to the
boundaries, the corner modes will be destroyed~\cite{Agarwala2020PRR}, suggesting the detrimental effects of the structural disorder on the higher-order
topological phases. However, we remarkably find that, in stark contrast to the case with chiral symmetry, the structural
disorder can in fact induce a higher-order topological phase transition from a topologically trivial phase in a crystalline geometry,
suggesting that the amorphous systems can favour the development of the second-order topology in 3D than
crystalline systems.
While the results are consistent with the previous finding of the on-site disorder induced first-order topological phase transitions,
such as topological Anderson insulators~\cite{Shen2009PRL,Beenakker2009PRL,Altland2015PRB,Zengyu2016SR},
we are the first to show that the structural disorder which connects a crystalline
material to an amorphous material can drive a topologically trivial phase to a higher-order nontrivial phase.
We finally generalize our results to the case with time-reversal symmetry (TRS) and show that
the second-order topological phase can exist in amorphous systems with
time-reversal symmetry characterized by a $\mathbb{Z}_2$ topological invariant, a spin winding number of the quadrupole moment and
a quantized longitudinal conductance of $4e^2/h$.

\emph{Model Hamiltonian.}---
To demonstrate the existence of amorphous SOTIs in 3D and
structural disorder induced SOTIs, we will work with the following tight-binding
Hamiltonian on 3D lattices with four degrees of freedom per site
\begin{equation}\label{Ham1}
\hat{H}_c=\sum_{\bf r} [M\hat{c}_{\bf r}^\dagger\tau_z\sigma_0 \hat{c}_{\bf r} +
\sum_{\bf d} t(|{\bf d}|)\hat{c}_{{\bf r}+{\bf d}}^\dagger T_c(\hat{{\bf d}}) \hat{c}_{\bf r}],
\end{equation}
where $\hat{c}_{\bf r}^\dagger=(\hat{c}_{{\bf r},1}^\dagger,\hat{c}_{{\bf r},2}^\dagger,
\hat{c}_{{\bf r},3}^\dagger,\hat{c}_{{\bf r},4}^\dagger)$
with $\hat{c}_{{\bf r},\alpha}^\dagger$ creating an electron of the $\alpha$th component at the site of position ${\bf r}$.
$\{\tau_{\nu}\}$ and $\{\sigma_{\nu}\}$ with $\nu=x,y,z$ are two sets of Pauli matrices acting on internal degrees of freedom.
The above Hamiltonian includes mass terms $M\tau_z\sigma_0$ at each site and
hopping terms between different sites.
For two sites at ${\bf r}$ and ${\bf r}+{\bf d}$, the hopping matrix
$T_c(\hat{\bf d})=[t_0\tau_z\sigma_0 + i t_1\tau_x(\hat{\bf d}\cdot \bm{\sigma}) + t_2(\hat{d}_x^2-\hat{d}_y^2)\tau_y\sigma_0]/2$
with ${\hat{{\bf d}}}={\bf d}/|{\bf d}|=(\hat{d}_x,\hat{d}_y,\hat{d}_z)$ being the unit vector along ${\bf d}$.
The hopping strength $t(|{\bf d}|)$ is chosen to decay exponentially with the distance as
$t(|{\bf d}|)=\Theta(d_c-|{\bf d}|)e^{-\lambda(|{\bf d}|/a-1)}$ consistent with real material scenarios. Here,
$d_c$ in the step function is a cutoff distance such that hoppings for $|{\bf d}|>d_c$ are neglected,
and we set the unit length of the system $a=1$ for simplicity.
Here, we choose the Hamiltonian parameters $t_0=t_1=t_2=1$ as the units of energy,
and the parameters for the hopping strength as $\lambda=3$ and $d_c=2.5$.
The system is assumed to be half-filled with the Fermi level at zero energy.

For a regular cubic lattice including only the nearest-neighbor hoppings,
the Hamiltonian (\ref{Ham1}) reduces to the paradigmatic model for
3D SOTIs hosting gapless chiral states localized at the hinges,
giving rise to a quantized longitudinal conductance of $2e^2/h$
along $z$.
The chiral hinge states are characterized by the Chern-Simons invariant,
which is protected to be quantized by a crystalline symmetry,
the combination of the time-reversal $\hat{T}$ and four-fold rotational $\hat{C}_4$ symmetry about the $z$ axis~\cite{Bernevig2018SciAdv}.
The requirement of the crystalline symmetry may suggest the absence of SOTIs in 3D amorphous systems.
Yet, besides the quantized Chern-Simons invariant, the topological phase can also
be protected by the winding number of the quadrupole moment about $k_z$ in momentum space,
reminiscent of a Chern insulator protected by the winding number of the Berry phase.
This makes it possible that SOTIs can exist in 3D amorphous materials without any spatial symmetry.
Indeed, we find a second-order topological insulating phase in a 3D amorphous system hosting
in-gap hinge states (see Fig.~\ref{Fig1}(a)) protected by the nontrivial winding number of
the quadrupole moment with respect to an inserted flux (see Fig.~\ref{Fig1}(b)). We further show that
structural disorder can induce a higher-order topological insulator in 3D.

\begin{figure}[t]
\includegraphics[width=3.4in]{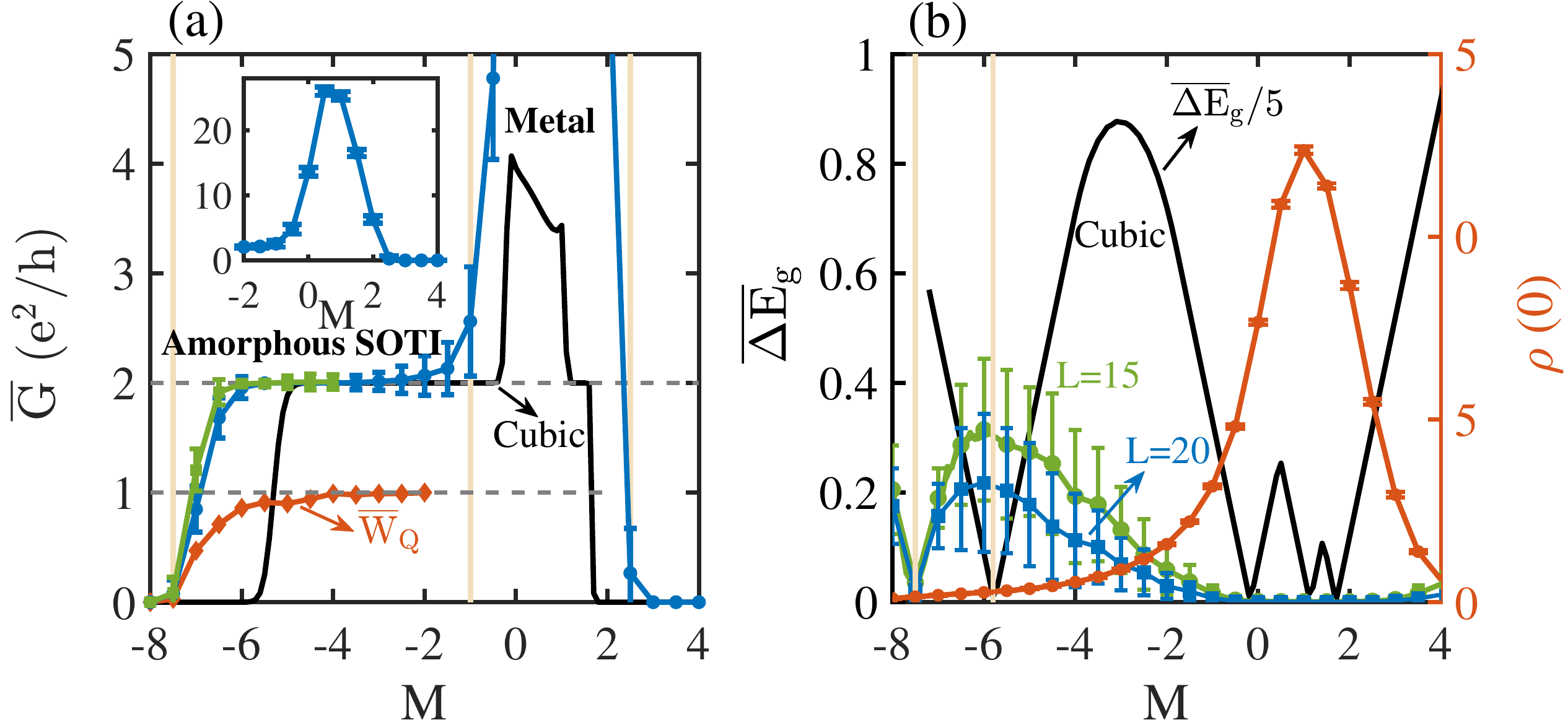}
\caption{(Color online)
(a) Configuration averaged longitudinal conductances $\overline{G}$ along $z$ (blue and green lines) and winding numbers $\overline{W}_Q$ of the quadrupole moment (red line) versus the mass $M$ for Hamiltonian (\ref{Ham1}) on amorphous lattices in comparison with the conductance for a cubic system (black line).
The blue, green and red lines correspond to systems with size $L=30,40, 16$, respectively.
Three distinct phases including amorphous SOTIs, metals and trivial insulators are identified for amorphous systems as separated by
the light yellow lines.
For the metal phase, the zoomed-in view of the conductance is plotted in the inset.
See the Supplementary Materials for standard deviations of the winding number.
(b) Configuration averaged bulk energy gaps (left vertical axis) versus $M$ for amorphous lattices
in comparison with the result for a cubic lattice
and the density of states (DOS) at zero energy $\rho(E=0)$ (right vertical axis) versus $M$ for amorphous lattices with $L=30$
calculated by the kernel polynomial method (KPM) with the expansion order $N_c=2^9$.}
\label{Fig2}
\end{figure}

\emph{Amorphous SOTIs.}---
To show that SOTIs can exist in amorphous systems,
we study the topological properties of the Hamiltonian (\ref{Ham1}) on completely random lattices,
where $N$ lattice sites are randomly placed in a cubic box of size $L$.
The coordinates of each site, ${\bf r}_{\nu}$ with $\nu=x,y,z$,
are randomly sampled from the uniform distribution in the interval $[0,L]$.
We set the average site density $\rho=N/L^3=1$ without loss of generality,
and take the average over 100 random configurations with standard deviations in numerical calculations.
By studying transport and band properties as well as the topological invariant for the Hamiltonian (\ref{Ham1}) on amorphous lattices,
we map out the phase diagram with respect to the mass $M$ and identify three distinct phases
including amorphous SOTIs, metals and trivial insulators, as shown in Fig.~\ref{Fig2}(a).

Since the chiral hinge states contribute a quantized longitudinal conductance of $2e^2/h$ along $z$ in
crystalline lattices, we expect that the quantized conductance may arise in amorphous systems when it becomes second-order
topological.
To numerically determine the zero-temperature two-terminal conductance $G$, we use the Landauer formula
\begin{equation}
G=\frac{e^2}{h} T(E_F),
\end{equation}
where $T(E_F)$ is the transmission probability from one lead to the other with incident electron energy at the Fermi level
for a randomized system connected to two semi-infinite leads along $z$; the
transmission probability is calculated using the nonequilibrium Green's function method~\cite{DattaBook,Sun2007PRB}.

In Fig.~\ref{Fig2}(a), we plot the sample averaged conductance $\overline{G}$ as a function of $M$ for amorphous lattices (blue and green lines),
remarkably illustrating the existence of a topological regime with the quantized conductance $\overline{G}=2e^2/h$ for $-7.5 \lesssim M \lesssim -1$
corresponding to an amorphous SOTI phase. Specifically,
as $M$ is increased from $M<-8$ in the trivial insulating phase,
we see that the conductance $\overline{G}$ suddenly rises to nonzero values around $M\approx-7.5$
and then enters into the quantized regime with $\overline{G}=2e^2/h$.
The results for $L=30$ and $L=40$ are plotted to show that around the critical point,
the conductance tends to become quantized for a system with a larger size.
In fact, the critical point between the trivial phase and the amorphous SOTI phase corresponds to a
bulk energy gap closing as shown in Fig.~\ref{Fig2}(b) because our disorder system respects a $\hat{C}_4 \hat{T}$
symmetry on average. Without the average symmetry, the topological phase can change through
a surface energy gap closing~\cite{Supplement}. Near the critical point,
the energy gap is small so that a larger system is required to obtain a nonzero quantized conductance.

When we further raise $M$, the system exhibits large values of the conductance suggesting a metallic phase up to $M \approx 2.5$
(see the inset of Fig.~\ref{Fig2}(a)).
Indeed, the bulk energy gap vanishes in this regime associated with large density of states (DOS) as shown in Fig.~\ref{Fig2}(b).

Figure~\ref{Fig2} also remarkably demonstrates the existence of a regime for $-7.5 \lesssim M \lesssim -5.8$
where $\overline{G}=2e^2/h$ in a random glass geometry while $G=0$ in a regular geometry, implying that structural disorder
can induce a topological phase transition from a topologically trivial phase to a higher-order topological nontrivial
one. The phenomenon is further evidenced by the gap closing points for different geometries as shown in Fig.~\ref{Fig2}(b).
We will elaborate on the structural disorder induced topological phase transition in the next section.

To further show that the quantized conductance arises from the topological bulk property of the system,
we evaluate the winding number of the quadrupole moment with respect to an inserted flux~\cite{footnote} defined as
\begin{equation}\label{winding1}
W_Q=\int_{0}^{2\pi} d\Phi_z \frac{\partial Q_{xy}(\Phi_z)}{\partial \Phi_z},
\end{equation}
where $\Phi_z$ is the flux twisting the boundary condition along $z$ and
$Q_{xy}(\Phi_z)$ is the quadrupole moment in the $(x,y)$ plane for the 3D random lattice under the flux $\Phi_z$.
The flux is added by replacing the hopping strength $t(|{\bf d}|)$ from site ${\bf r}$
to site ${\bf r}+{\bf d}$ with $t(|{\bf d}|)e^{i\Phi_z d_z/L_z}$.
$Q_{xy}(\Phi_z)$ is calculated using occupied single-particle states $|\psi_n(\Phi_z)\rangle$ ($n=1,\cdots,N_{occ}$) of the Hamiltonian
under periodic boundary conditions as
$Q_{xy}(\Phi_z)=\frac{1}{2\pi}\text{Im}\log\det(U_Q(\Phi_z))$ where
$[U_{Q}(\Phi_z)]_{mn}=\langle \psi_m(\Phi_z)| \hat{U}_{Q} |\psi_n(\Phi_z)\rangle$
and $\hat{U}_{Q}=e^{i2\pi\hat{x}\hat{y}/(L_x L_y)}$ with $\hat{x}$ ($\hat{y}$) denoting the
$x$-position ($y$-position) operator for a single electron~\cite{Cho2019PRB,Wheeler2019PRB}.
As the flux $\Phi_z$ varies from $0$ to $2\pi$, the quadrupole moment $Q_{xy}(\Phi_z)$ should exhibit a nontrivial winding number for the SOTI phase.

In Fig.~\ref{Fig2}(a), we plot the calculated winding number $\overline{W}_Q$ averaged over configurations for amorphous lattices
with respect to $M$. We see that $\overline{W}_Q$ grows up rapidly
from zero to nonzero values as $M$ increases to a critical point
$M\approx-7.5$, and then becomes close to the quantized value $\overline{W}_Q=1$,
reflecting a phase transition from the topologically trivial phase to the amorphous SOTI phase,
in consistent with the results of the conductance.

\begin{figure}[t]
\includegraphics[width=3.4in]{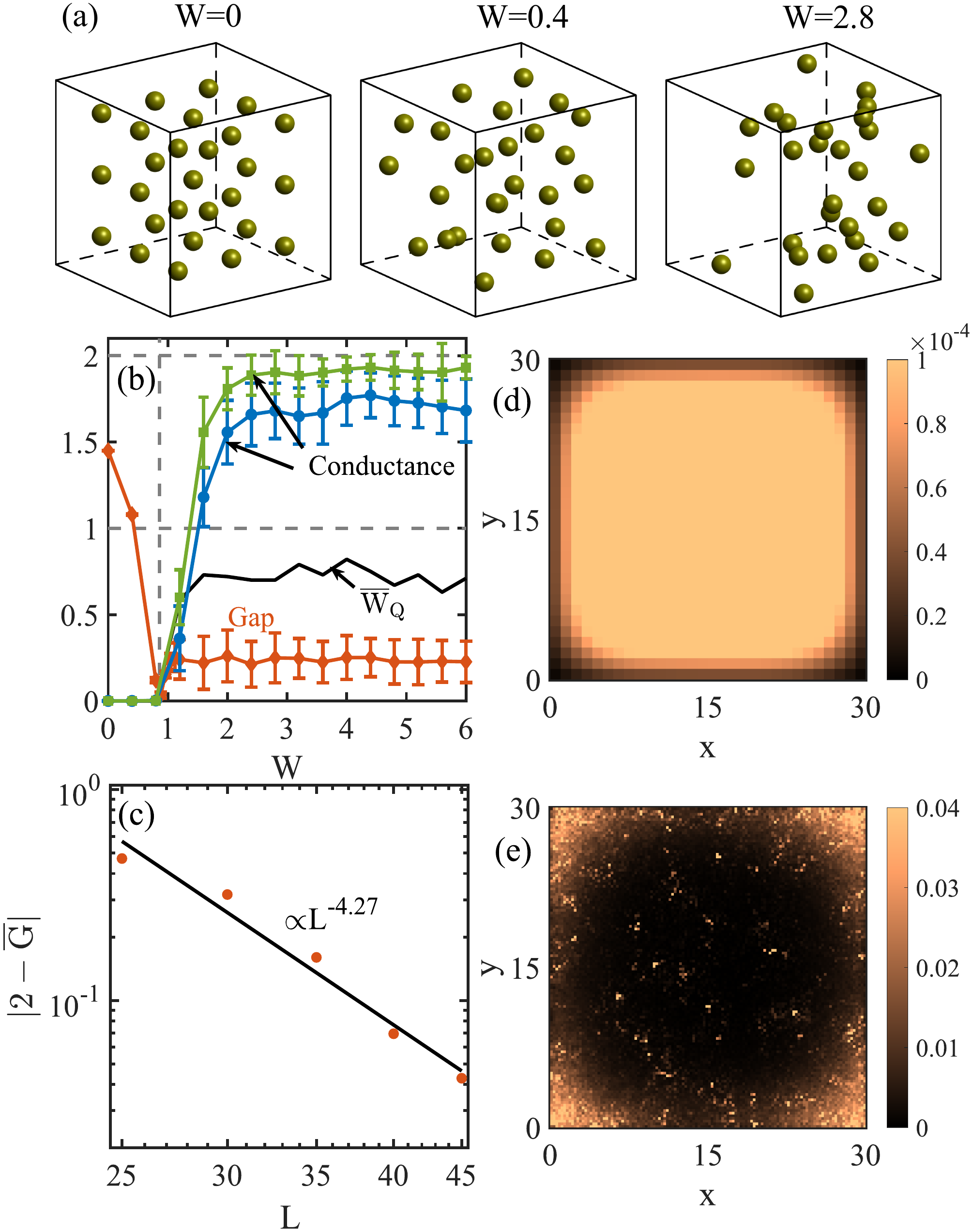}
\caption{(Color online)
(a) Schematics of lattice structures for three structural disorder strengths $W$ added on a regular cubic lattice.
(b) Configuration averaged longitudinal conductances along $z$, quadrupole moment winding numbers and bulk energy gaps
versus $W$.
(c) The finite-size scaling of the conductance versus the system size $L$ for $W=6$,
which displays a power-law decay fitted by a black line.
A top view of LDOS at zero energy for (d) $W=0$ and (e) $W=6$, obtained by averaging over different random configurations and
summing over the coordinates along $z$.
For (b-e), we take $M=-6.5$ in Hamiltonian (\ref{Ham1}).}
\label{Fig3}
\end{figure}

\emph{Structural disorder induced SOTIs.}---
To demonstrate the structural disorder induced topological phase transition, we consider adding structural disorder gradually on the cubic lattice as follows.
For each lattice site, we add a random displacement along three orthogonal directions from the corresponding regular position in the cubic lattice
based on the uniform distribution in the interval $[-{W}/{2},{W}/{2}]$,
where $W$ represents the strength of structural disorder (see Fig.~\ref{Fig3}(a) for typical configurations).
When $W$ is increased from zero, the lattice structure changes from a cubic lattice to a slightly irregular lattice
and then to a completely random lattice.

To see the structural disorder induced topological phase transition, in Fig.~\ref{Fig3}(b),
we plot the longitudinal conductance $\overline{G}$ along $z$ and the winding number $\overline{W}_Q$ of the quadrupole moment
with respect to the structural disorder strength $W$ for $M=-6.5$.
For small $W's$, the system deviates slightly from the cubic lattice and remains in the topologically trivial phase
with zero conductance and winding number.
As we further increase $W$, both $\overline{G}$ and $\overline{W}_Q$ suddenly rise to nonzero values at $W\approx 0.8$, indicating
that the system undergoes a topological phase transition entering into the SOTI phase.
The topological phase transition is also identified by the bulk energy gap closing at the critical point as shown in Fig.~\ref{Fig3}(b).
We note that both $\overline{G}$ and $\overline{W}_Q$ averaged over random configurations are not quantized due to finite-size effects.
To illustrate this, we calculate $|2-\overline{G}|$ in units of $e^2/h$ for different system sizes when $W=6$
and show that $|2-\overline{G}|\propto L^{-4.27}$ (see Fig.~\ref{Fig3}(c)), indicating that $\overline{G}$ will approach the quantized conductance of
$2e^2/h$ in the thermodynamic limit.

To further identify the existence of gapless hinge states in the structural disorder induced SOTI phase,
we compute the LDOS at zero energy
under open boundary conditions along $x$ and $y$ directions and periodic boundary conditions along $z$.
In Fig.~\ref{Fig3}(d) and (e), we display the LDOS summed over the coordinates along $z$ for $W=0$ and $W=6$, respectively.
The LDOS clearly shows the existence of in-gap states localized near the hinges when $W=6$ in the amorphous SOTI phase,
in stark contrast to the trivial phase without the hinge states when $W=0$.

\emph{Amorphous SOTIs with TRS.}---
We now construct a model for SOTIs with TRS on random lattices with eight degrees of freedom per site
\begin{equation}\label{Ham2}
\hat{H}_h=\sum_{{\bf r}} [M\hat{c}_{{\bf r}}^\dagger \tau_z s_0 \sigma_0 \hat{c}_{{\bf r}}
+\sum_{{\bf d}} t(|{\bf d}|)\hat{c}_{{\bf r}+{\bf d}}^\dagger T_h(\hat{{\bf d}}) \hat{c}_{{\bf r}}],
\end{equation}
where $\hat{c}_{{\bf r}}^\dagger=(\hat{c}_{{\bf r},1}^\dagger,\cdots,\hat{c}_{{\bf r},8}^\dagger)$ with $\hat{c}_{{\bf r},\alpha}^\dagger$ being
a creation operator for an electron of the $\alpha$th component at site ${\bf r}$.
Besides $\{\tau_{\nu}\}$ and $\{\sigma_{\nu}\}$, $\{s_{\nu}\}$ with $\nu=x,y,z$ is also a set of Pauli matrices.
Here, the hopping matrix between two different sites ${\bf r}$ and ${\bf r}+{\bf d}$ is described by
$T_h(\hat{\bf d})=[t_0\tau_z s_0\sigma_0 + i t_1\tau_x s_0(\hat{\bf d}\cdot \bm{\sigma}) + t_2(\hat{d}_x^2-\hat{d}_y^2)\tau_y s_y\sigma_0
+i t_3 \hat{d}_z \tau_y s_x\sigma_0]/2$.
The Hamiltonian $\hat{H}_h$ respects the TRS $\hat{T}$ implemented through
$\hat{T}i\hat{T}^{-1}=-i$ and $\hat{T}\hat{c}_{\bf r}\hat{T}^{-1}=U_T\hat{c}_{\bf r}$ with $U_T=i\tau_0 s_0\sigma_y$,
such that $\hat{T}\hat{H}_h\hat{T}^{-1}=\hat{H}_h$.
When $t_3=0$, $\hat{H}_h$ respects an additional U(1) pseudospin rotational symmetry with the conservation of the pseudospin
$s_y$ so that $\hat{H}_h$ can be written as the direct sum of two copies of Hamiltonian (\ref{Ham1}) with
opposite signs of $t_2$ due to opposite eigenvalues of the Pauli matrix $s_y$.
In this case, we introduce a spin quadrupole moment winding number to characterize the helical hinge states present
in an amorphous SOTI with TRS~\cite{Supplement}. For nonzero $t_3$, while $s_y$ is no longer conserved, we find that the spin winding number can still characterize the amorphous SOTI with TRS when $t_3$ is not very large~\cite{Supplement}.

\begin{figure}[t]
\includegraphics[width=1.8in]{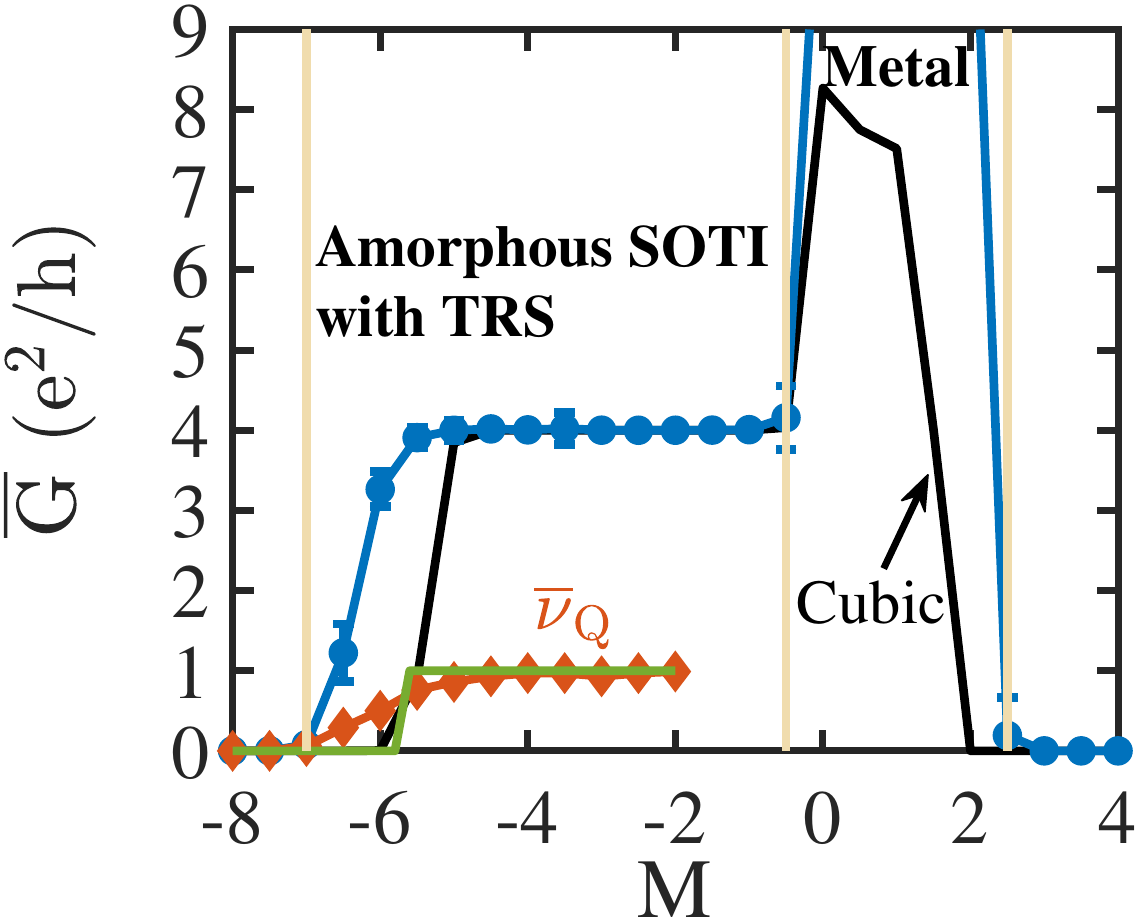}
\caption{(Color online)
Sample averaged longitudinal conductances $\overline{G}$ along $z$ (blue line) and $\mathbb{Z}_2$ invariants $\overline{\nu}_Q$ (red line) versus the mass $M$ for Hamiltonian (\ref{Ham2}) on amorphous lattices in comparison with the conductance (black line) and the $\mathbb{Z}_2$ invariant (green line) on a cubic lattice.
The blue (red) line corresponds to a system with size $L=30$ ($L=12$).
}
\label{Fig4}
\end{figure}

In a generic case with TRS,
we further derive a $\mathbb{Z}_2$ topological invariant $\nu_Q \in \{0,1\}$ defined as
\begin{equation}
(-1)^{\nu_Q}
=\frac{\textrm{Pf}[A(\pi)]}{\textrm{Pf}[A(0)]} \sqrt{\frac{\det[A(0)]}{\det[A(\pi)]}},
\end{equation}
where $\sqrt{\frac{\det[A(0)]}{\det[A(\pi)]}}=\exp \left\{ -\frac{1}{2}\int_{0}^{\pi} d\Phi_z \frac{\partial \log \det[A(\Phi_z)]}{\partial \Phi_z} \right\}$, $\textrm{Pf}[\cdot]$ represents the Pfaffian of an antisymmetric matrix, and the matrix $A(\Phi_z)$ is defined as
$
[A(\Phi_z)]_{mn}=\langle \psi_m(-\Phi_z)| \hat{U}_{Q} {T} |\psi_n(\Phi_z)\rangle
$
for occupied single-particle states $|\psi_n(\Phi_z)\rangle$ [$|\psi_m(-\Phi_z)\rangle$] of Hamiltonian (\ref{Ham2}) with the flux $\Phi_z$ [$-\Phi_z$]. For numerical calculations, we derive a simplified formulation for the invariant based on the quadrupole moment~\cite{Supplement}.

In Fig.~\ref{Fig4},
we map out the phase diagram for Hamiltonian (\ref{Ham2}) on random lattices.
The existence of quantized conductances of $4e^2/h$ and $\mathbb{Z}_2$ invariants
indicate the presence of amorphous SOTIs with TRS. Apart from the topological insulating phase,
a metal and a trivial insulator are also identified.

In summary, we have demonstrated the existence of a SOTI in an amorphous system and
predicted a structural disorder induced topological phase transition from a topologically trivial phase
in a crystalline lattice to a SOTI. Our results should be far more generic than our model given that
our analysis indicates that the SOTI does not require any spatial order.
Our results also have important implications that amorphous solids
may broadly support the SOTI phase.
Specifically, the bismuth crystal has been experimentally identified as a 3D SOTI with helical hinge states~\cite{Neupert2018NP}.
We thus expect that the 3D amorphous SOTIs may be observed in amorphous bismuth.
In fact, 3D amorphous topological insulators with TRS has been experimentally observed in the
films of Bi$_2$Se$_3$ grown on the amorphous substrates~\cite{Corbae2019exp}. We expect that amorphous bismuth can be fabricated similarly.
In addition, amorphous SOTIs and the structural disorder induced topological phase transition can also be experimentally
observed in metamaterials,
such as photonic, phononic and electric circuit systems~\cite{Huber2018Nature,Bahl2018Nature,Thomale2018NP}.

\begin{acknowledgments}
We thank Y.-L. Tao for helpful discussion. The work is
supported by the National Natural Science Foundation
of China (Grant No. 11974201), the start-up fund from Tsinghua University, and
the National Thousand-Young-Talents Program.
\end{acknowledgments}

\begin{widetext}
\setcounter{equation}{0} \setcounter{figure}{0} \setcounter{table}{0} %
\renewcommand{\theequation}{S\arabic{equation}} \renewcommand{\thefigure}{S%
\arabic{figure}} \renewcommand{\bibnumfmt}[1]{[S#1]} 

In the supplementary material, we will provide the winding number with a standard deviation in Section S-1,
discuss the effects of the $\hat{C}_4\hat{T}$ symmetry on average in Section S-2,
deduce a $\mathbb{Z}_2$ topological invariant for a SOTI with TRS in Section S-3,
introduce a spin quadrupole moment winding number in Section S-4, and finally
present the relation between the $\mathbb{Z}_2$ invariant and the spin quadrupole moment winding number in Section S-5.

\section{S-1. The winding number with a standard deviation}
In this section, we show the fluctuations of the winding number of the quadrupole moment due to
limited system sizes by plotting their standard deviations in Fig.~\ref{figS1}.
Since the winding numbers can only take integer values for distinct disorder realizations
instead of exhibiting a Gaussian distribution,
we also display the number of disorder samples with $W_Q=1$ in 100 disorder realizations, illustrating
the percentage of the total realizations that yields $W_Q=1$.

\begin{figure}[t]
\includegraphics[width=7in]{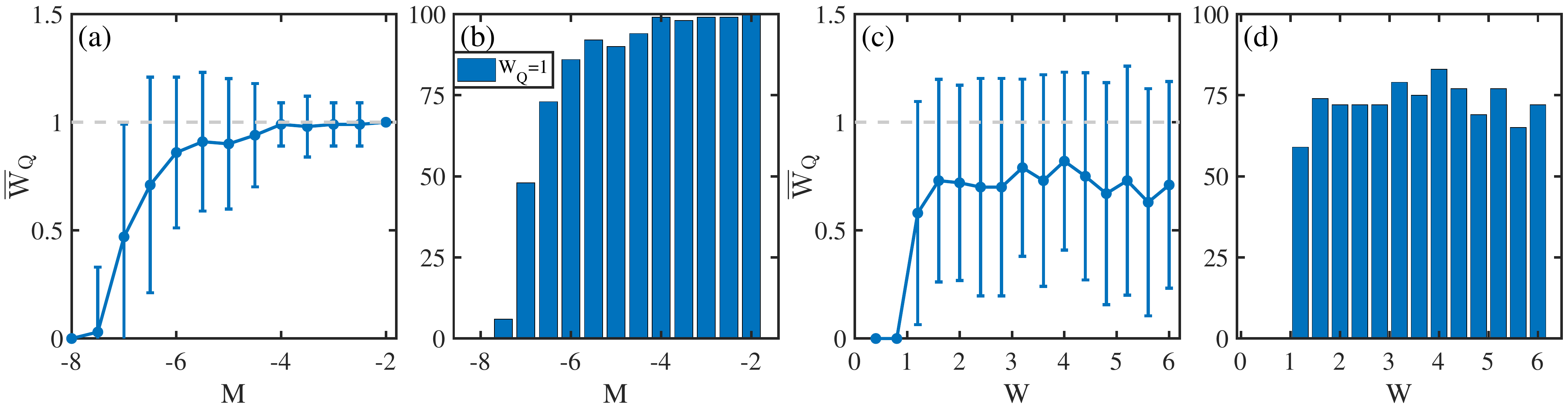}
\caption{(Color online)
Configuration averaged winding numbers of the quadrupole moment $\overline{W}_Q$ with a standard deviation
for (a) versus $M$ and (c) versus $W$.
The number of configurations with $W_Q=1$ in 100 disorder realizations for (b) versus $M$ and (d) versus $W$.
In (a) and (b) [(c) and (d)], the system parameters are the same as those in Fig. 2(a) [in Fig. 3(b)] in the main text.
}
\label{figS1}
\end{figure}

\section{S-2. The $\hat{C}_4\hat{T}$ symmetry on average}
In this section, we will discuss the effect of the average $\hat{C}_4\hat{T}$ symmetry on topological phase transitions.
The main conclusion is that in the presence of the
average symmetry, the system undergoes a topological phase transition from a trivial to a nontrivial higher-order topological phase
through a bulk energy gap closing, while in the absence of
the average symmetry, it can occur through a surface energy gap closing.

Let us first consider the Hamiltonian (1) in the main text and define the $\hat{C}_4\hat{T}$ symmetry on average.
For an individual disorder realization, the system Hamiltonian $\hat{H}_c$ clearly breaks the $\hat{C}_4\hat{T}$ symmetry due to its random geometry configuration, i.e., $\hat{C}_4\hat{T}\hat{H}_c(\hat{C}_4\hat{T})^{-1}\neq \hat{H}_c$. Specifically,
\begin{eqnarray}
\hat{C}_4\hat{T}\hat{H}_c(\hat{C}_4\hat{T})^{-1}
&=&\sum_{{\bf r}\in S} \left[ M\hat{c}_{D_{\hat{C}_4}{\bf r}}^\dagger \tau_z\sigma_0 \hat{c}_{D_{\hat{C}_4}{\bf r}}
+\sum_{\substack{{\bf d}={\bf r}_1-{\bf r} \\ {\bf r}_1 \in S, {\bf r}_1\neq {\bf r} }} t(|d|)\hat{c}_{D_{\hat{C}_4}({\bf r}+{\bf d})}^\dagger T_c(D_{\hat{C}_4}\hat{\bf d}) \hat{c}_{D_{\hat{C}_4}{\bf r}} \right] \\
&=&\sum_{{\bf r}^\prime \in S^\prime } \left[ M\hat{c}_{{\bf r}^\prime}^\dagger \tau_z\sigma_0 \hat{c}_{{\bf r}^\prime}
+\sum_{\substack{{\bf d}^\prime={\bf r}_1^\prime-{\bf r}^\prime \\ {\bf r}_1^\prime \in S^\prime, {\bf r}_1^\prime \neq {\bf r}^\prime }} t(|d^\prime|)\hat{c}_{{\bf r}^\prime+{\bf d}^\prime}^\dagger T_c(\hat{\bf d}^\prime) \hat{c}_{{\bf r}^\prime} \right],
\end{eqnarray}
where we have used the results implemented through the symmetry operation: $\hat{C}_4\hat{T} \hat{c}_{\bf r}(\hat{C}_4\hat{T})^{-1}=i\tau_0\sigma_y e^{-i\frac{\pi}{4}\sigma_z}\hat{c}_{D_{\hat{C}_4}{\bf r}}$ and
$\hat{C}_4\hat{T}i(\hat{C}_4\hat{T})^{-1}=-i$. Here, $S$ is a set consisting of position vectors of all sites in a configuration,
and
$S^\prime=D_{\hat{C}_4} S\equiv \{ D_{\hat{C}_4}{\bf r}: {\bf r}\in S\}$ is a set obtained by rotating all position vectors in $S$
with $D_{\hat{C}_4}{\bf r}=(-y,x,z)$ that rotates a vector $\bf r$ in a counterclockwise direction about $z$ by 90 degrees.
For a cubic lattice configuration, since $S^\prime = S$, we have $\hat{C}_4\hat{T}\hat{H}_c(\hat{C}_4\hat{T})^{-1}= \hat{H}_c$
so that the system respects
the $\hat{C}_4\hat{T}$ symmetry. In contrast, for a typical sample with randomly distributed sites, $S^\prime \neq S$ and thus $\hat{C}_4\hat{T}\hat{H}_c(\hat{C}_4\hat{T})^{-1}\neq \hat{H}_c$, indicating that $\hat{H}_c$ as a single system does not respect the $\hat{C}_4\hat{T}$ symmetry.
We now consider all systems in a statistical ensemble. If a Hamiltonian
and its symmetry conjugate partner appear in the ensemble with the same
probability, then the ensemble respects a symmetry on average~\cite{FuliangPRL}.
In our case, if there is a configuration $S$ in a statistical ensemble, then one can always find a configuration $S^\prime=D_{\hat{C}_4} S$ that appears with the same probability as $S$.
This indicates that $\hat{H}_c$ and its $\hat{C}_4\hat{T}$ conjugate partner $\hat{C}_4\hat{T}\hat{H}_c(\hat{C}_4\hat{T})^{-1}$ exist in an ensemble with the same probability, and hence our system respects the
$\hat{C}_4\hat{T}$ symmetry on average.

With this average symmetry, we argue that a single system in an ensemble must close its bulk energy gap in order to
change its higher-order topological property. Specifically, consider a subsystem $\hat{H}_1$ with its $z$ coordinate in the interval $[z_0,z_0+\Delta z]$ in a large system as shown in Fig.~\ref{figS2}.
If we view $\hat{H}_1$ as a single system with spatial configuration $S_1$, then chiral hinges states in $\hat{H}_1$ are not protected by a bulk energy gap since $\hat{H}_1$ does not respect the $\hat{C}_4\hat{T}$ symmetry.
Instead, they can appear through a surface energy gap closing in $\hat{H}_1$ associated with the change of the winding number of the quadrupole moment. Without loss of generality, we suppose that the energy gap on the $x$-normal surface closes.
Consider a very large system (e.g., infinitely long along $z$),
we expect that the spatial configuration $S_2=D_{\hat{C}_4} S_1+\alpha {\bf e}_z$
obtained by rotating $S_1$ about $z$ by 90 degrees and shifting its $z$ coordinate by $\alpha$
should always exist.
Let $\hat{H}_2$ be another subsystem with the spatial configuration $S_2$ as shown in Fig.~\ref{figS2}. Clearly, $\hat{H}_2$ is the $\hat{C}_4\hat{T}$ conjugate partner of $\hat{H}_1$ with all its $z$ coordinates shifting by $\alpha$, which does not have any effects. Since the energy gap on the $x$-normal surface for $\hat{H}_1$ closes, the energy gap on the $y$-normal surface for $\hat{H}_2$ must close.
This indicates that as a whole system including both $\hat{H}_1$ and $\hat{H}_2$, the system must close its energy gap on both $x$-normal and $y$-normal surfaces, suggesting that its bulk energy gap should vanish. Therefore, we conclude that the chiral hinge modes are protected by a bulk energy gap if a system respects a $\hat{C}_4\hat{T}$ symmetry on average. Indeed, for a system with the average symmetry,
we observe the bulk energy gap closing when the system changes from a topologically trivial phase to a nontrivial one in Fig. 2(b) in the main text.

\begin{figure}[t]
\includegraphics[width=2.5in]{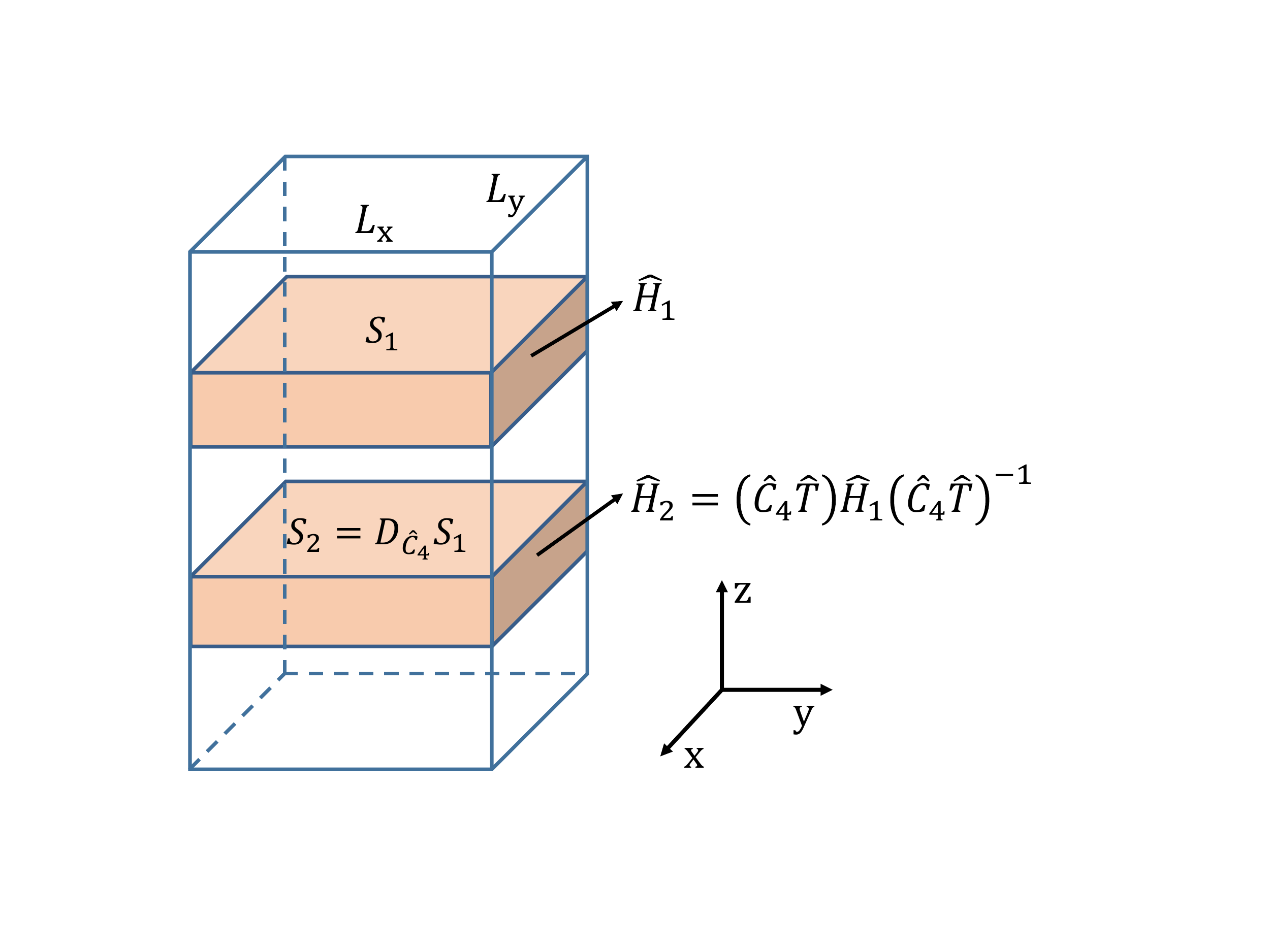}
\caption{(Color online)
Illustration of a system including two subsystems $\hat{H}_1$ and $\hat{H}_2$ corresponding to spatial configurations of $S_1$
and $S_2$, respectively.
}
\label{figS2}
\end{figure}

\begin{figure}[t]
\includegraphics[width=5.5in]{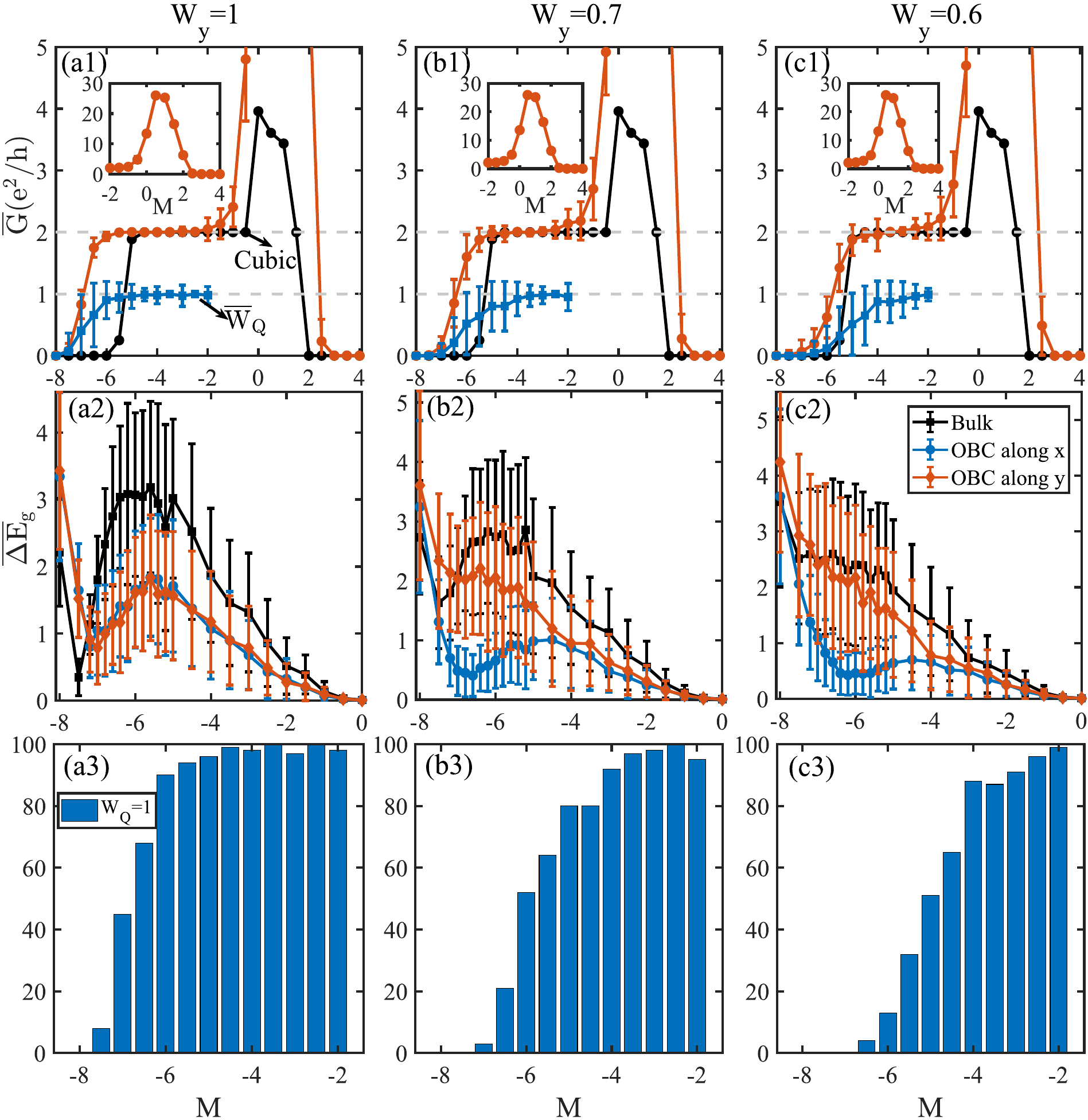}
\caption{(Color online)
(a1)-(c1) Configuration averaged two-terminal longitudinal conductances $\overline{G}$ (system size $L=30$) and winding numbers of the quadrupole moment $\overline{W}_Q$ (system size $L=16$) with respect to the mass $M$.
The black lines show the results for a cubic lattice configuration.
(a2)-(c2) Configuration averaged bulk and surface energy gaps (system size $L=15$) with respect to the mass $M$.
The black, blue and red lines correspond to a system with periodic boundaries along all directions, a system with open boundaries along $x$
and periodic boundaries along other directions, and a system with open boundaries along $y$ and periodic boundaries along other directions, respectively.
(a3)-(c3) The number of samples with $W_Q=1$ in 100 disorder realizations versus the mass $M$.
Here, in (a1)-(a3) $W_y=1$, in (b1)-(b3) $W_y=0.7$, and in (c1)-(c3) $W_y=0.6$.
}
\label{figS3}
\end{figure}

But this does not mean that chiral hinge states cannot exist in an amorphous system when the average $\hat{C}_4\hat{T}$ symmetry is broken.
In fact, we find their existence in the absence of the average symmetry.
To lift the average $\hat{C}_4\hat{T}$ symmetry, we consider a position ensemble $\Xi$ comprised of all allowable spatial configurations of atom sites.
For $\forall S \in \Xi$, if another element $S^\prime=D_{\hat{C}_4}S$ appears with the same probability as $S$ in $\Xi$,
then we say that the ensemble respects a statistical $C_4$ symmetry.
To break the symmetry, we enforce a constraint that $y$ coordinates are not as random as $x$ coordinates.
As a result, $S$ and $S^\prime$ do not appear in $\Xi$ with equal probabilities so that
$\hat{H}_c$ and its $\hat{C}_4\hat{T}$ conjugate partner $\hat{C}_4\hat{T}\hat{H}_c(\hat{C}_4\hat{T})^{-1}$ do not emerge with the same probability, lifting the $\hat{C}_4\hat{T}$ symmetry on average.

To be concrete, we obtain atom position configurations by taking ${\bf r}=[{\bf R}+(W_x \Delta_x,W_y \Delta_y,W_z \Delta_z )] \mod L$ where $\bf R$ is the site position vector in a cubic lattice with system size of $L$, and $\Delta_x$, $\Delta_y$ and $\Delta_z$ are randomly sampled from the uniform
distribution in the interval $[-0.5,0.5]$. To lift the average $\hat{C}_4\hat{T}$ symmetry, we set $W_y=W_z=L$ and $W_x=0.6, 0.7, 1.0$.
For all these system parameters, we find the existence of amorphous SOTIs illustrated
by sample averaged longitudinal conductances and winding numbers of the quadrupole moment (see Fig.~\ref{figS3}).
One can also see that the topological regions become smaller when the break of the symmetry becomes stronger,
which is also illustrated by the shifting of the position of the minimum energy gap toward the right side as
$W_y$ decreases.
Interestingly, the figure also suggests that when the symmetry is slightly broken for $W_y=1$, the topological property
changes through a bulk energy gap closing, while when the break becomes stronger for $W_y=0.7,0.6$,
the topological phase changes through a surface energy gap closing. We therefore conclude that without the
$\hat{C}_4\hat{T}$ symmetry on average, the topological phase may change either through a bulk energy gap
closing or through a surface energy gap closing depending on the destruction level of the symmetry.

\section{S-3. A $\mathbb{Z}_2$ topological invariant for SOTIs with TRS}
\subsection{A. Theoretical deduction}
In this subsection, we give a detailed derivation of a $\mathbb{Z}_2$ topological invariant based on the quadrupole moment for a noninteracting
helical SOTI with TRS in 3D, which has been introduced in the main text. To be concrete,
we consider a Hamiltonian $\hat{H}_h$ for free electrons with TRS $\hat{T}$ satisfying
\begin{equation}
\hat{T}\hat{H}_h\hat{T}^{-1}=\hat{H}_h
\end{equation}
and $\hat{T}^2=-1$ so that the system belongs to the class AII
according to Altland-Zirnbauer (AZ) classification~\cite{Altland1997PRB,Chiu2016RMP}.

For a disordered system, we introduce a flux $\Phi_z$ twisting the boundary conditions along $z$ by adding Peierls phase
factors to hopping amplitudes in the Hamiltonian $\hat{H}_h$. Under the time-reversal transformation $\hat{T}$,
$\hat{H}_h(\Phi_z)$ transforms as
\begin{equation}
\hat{T}\hat{H}_h(\Phi_z)\hat{T}^{-1}=\hat{H}_h(-\Phi_z),
\end{equation}
where the time-reversal operator reverses the sign of the flux $\Phi_z$ due to the complex conjugation acting on Peierls phase
factors in $\hat{H}_h(\Phi_z)$.
We remark that the following derivations for the topological invariant can also be applied to periodic systems with translational symmetries
where the flux $\Phi_z$ is replaced with quasi-momentum $k_z$ in momentum space.
Since we focus on the noninteracting case, for clarity, we write the Hamiltonian as
\begin{equation}
\hat{H}=\hat{\Psi}^\dagger [H_h] \hat{\Psi},
\end{equation}
where $\hat{\Psi}^\dagger$ is a row vector comprised of
$\hat{c}_{{\bf r}\alpha}$, and $[H_h]$ is the matrix representation of the first quantization Hamiltonian $H_h$. In the first quantization form,
the Hamiltonian transforms through the time-reversal operation as
\begin{equation}
{T}{H_h}(\Phi_z){T}^{-1}={H_h}(-\Phi_z),
\end{equation}
where $T=-i \sigma_y \kappa$ with $\kappa$ being the complex conjugate operator.

We now define a $\Phi_z$-dependent matrix $A(\Phi_z)$, which is related to the quadrupole moment as
\begin{equation}\label{A_phiz}
[A(\Phi_z)]_{mn}=\langle \psi_m(-\Phi_z)| \hat{U}_{Q} {T} |\psi_n(\Phi_z)\rangle,
\end{equation}
where $\hat{U}_{Q}=e^{i2\pi\hat{x}\hat{y}/(L_x L_y)}$ with $\hat{x}$ ($\hat{y}$) denoting the $x$-position ($y$-position) operator for a single electron, and
$|\psi_n(\Phi_z)\rangle$ is the $n$th occupied single-particle eigenstate of the Hamiltonian ${H_h}(\Phi_z)$.
For each $\Phi_z$, $A(\Phi_z)$ is an $N_{occ}\times N_{occ}$ matrix with $N_{occ}$ being the total number of occupied states.
In the following, we will prove the following two properties that $A(\Phi_z)$ satisfies:
\begin{align}
&A(\Phi_z)=A(\Phi_z+2\pi), \\
&A(\Phi_z)=-[A(-\Phi_z)]^{T},
\end{align}
where $[A(\Phi_z)]^{T}$ denotes the transpose of the matrix $A(\Phi_z)$.

To prove the $2\pi$-periodicity of $A(\Phi_z)$ about the flux $\Phi_z$,
we rely on a fact that ${H_h}(\Phi_z+2\pi)$ and ${H_h}(\Phi_z)$ are related by a unitary transformation
\begin{equation}
{H_h}(\Phi_z+2\pi)=e^{i2\pi\hat{z}/L_z}{H_h}(\Phi_z)e^{-i2\pi\hat{z}/L_z}.
\end{equation}
We thus can choose the eigenstates $\{|\psi_n(\Phi_z+2\pi)\rangle\}$ of ${H}_h(\Phi_z+2\pi)$ to be $|\psi_n(\Phi_z+2\pi)\rangle=e^{i2\pi\hat{z}/L_z}|\psi_n(\Phi_z)\rangle$.
As a result, we have
\begin{align}
[A(\Phi_z+2\pi)]_{mn}&=\langle \psi_m(-\Phi_z-2\pi)| \hat{U}_{Q} {T} |\psi_n(\Phi_z+2\pi)\rangle \nonumber \\
&=\langle \psi_m(-\Phi_z)|e^{i2\pi\hat{z}/L_z} \hat{U}_{Q} {T} e^{i2\pi\hat{z}/L_z}|\psi_n(\Phi_z)\rangle \nonumber \\
&=\langle \psi_m(-\Phi_z)| \hat{U}_{Q} {T} e^{-i2\pi\hat{z}/L_z} e^{i2\pi\hat{z}/L_z}|\psi_n(\Phi_z)\rangle \nonumber \\
&=\langle \psi_m(-\Phi_z)| \hat{U}_{Q} {T} |\psi_n(\Phi_z)\rangle \nonumber \\
&=[A(\Phi_z)]_{mn},
\end{align}
where we have used the anti-unitary property of the time-reversal operator ${T}$ in the derivation.
We thus obtain the relation that $A(\Phi_z)=A(\Phi_z+2\pi)$.

To prove the antisymmetric property of $A(\Phi_z)$, we rely on the anti-unitary property of ${T}$ and ${T}^2=-1$.
Specifically,
\begin{align}
[A(\Phi_z)]_{mn}&=\langle \psi_m(-\Phi_z)| \hat{U}_{Q} {T} |\psi_n(\Phi_z)\rangle \nonumber \\
&=\langle {T}\psi_m(-\Phi_z)| {T} \hat{U}_{Q} {T} |\psi_n(\Phi_z)\rangle^{*} \nonumber \\
&=\langle {T}\psi_m(-\Phi_z)| \hat{U}_{Q}^{*} {T} {T} |\psi_n(\Phi_z)\rangle^{*} \nonumber \\
&=- \langle {T}\psi_m(-\Phi_z)| \hat{U}_{Q}^{\dagger} |\psi_n(\Phi_z)\rangle^{*} \nonumber \\
&=- \langle \psi_n(\Phi_z)| \hat{U}_{Q} {T}|\psi_m(-\Phi_z)\rangle \nonumber \\
&=- [A(-\Phi_z)]_{nm}.
\end{align}
We thus obtain the relation that $A(\Phi_z)=-[A(-\Phi_z)]^{T}$.
With the aid of $A(\Phi_z)=A(\Phi_z+2\pi)$, we further obtain that the matrix $A(\Phi_z)$ is antisymmetric at two time-reversal symmetric points $\Phi_z=0$ and $\Phi_z=\pi$.

We now proceed to construct a Hermitian Hamiltonian matrix based on $A(\Phi_z)$ as
\begin{equation}
{H}_{A}(\Phi_z)=
\begin{pmatrix}
  0 & A(\Phi_z) \\
  [A(\Phi_z)]^\dagger & 0 \\
\end{pmatrix}.
\end{equation}
Here, we assume that the matrix $A(\Phi_z)$ is invertible so that ${H}_{A}(\Phi_z)$ has a gap at zero energy.
Since ${H}_{A}(\Phi_z)$ is $2\pi$-periodic about $\Phi_z$, we can view ${H}_{A}(\Phi_z)$
as a one-dimensional (1D) gapped Hamiltonian in momentum space with $\Phi_z$ being the quasi-momentum.
${H}_{A}$ thus respects the chiral symmetry, the time-reversal symmetry and the particle-hole symmetry,
since it satisfies the following symmetry constraints,
\begin{align}
&{S}_{A} {H}_{A}(\Phi_z) {S}_{A}^{-1} = -{H}_{A}(\Phi_z), \\
&{T}_{A} {H}_{A}(\Phi_z) {T}_{A}^{-1} = {H}_{A}(-\Phi_z), \\
&{P}_{A} {H}_{A}(\Phi_z) {P}_{A}^{-1} = -{H}_{A}(-\Phi_z),
\end{align}
where
${S}_{A}=
\begin{pmatrix}
  {I}_{N_{occ}} & 0 \\
  0 & -{I}_{N_{occ}} \\
\end{pmatrix}
$ with ${I}_{N_{occ}}$ being an $N_{occ} \times N_{occ}$ identity matrix, the time-reversal operator
${T}_{A}=
\begin{pmatrix}
  0 & -{I}_{N_{occ}} \\
  {I}_{N_{occ}} & 0 \\
\end{pmatrix}\kappa
$ with ${T}_{A}^2=-1$, and the particle-hole operator
${P}_{A}=
\begin{pmatrix}
  0 & {I}_{N_{occ}} \\
  {I}_{N_{occ}} & 0 \\
\end{pmatrix}\kappa
$ with ${P}_{A}^2=1$.
Due to the above symmetry constraints,
${H}_{A}(\Phi_z)$ belongs to the class DIII for 1D free electron systems and the bulk topology is $\mathbb{Z}_2$ classified according to the AZ classification~\cite{Chiu2016RMP}.
We thus use the $\mathbb{Z}_2$ topological invariant of ${H}_{A}(\Phi_z)$ to classify a generic 3D
SOTI with TRS discussed in the main text.

For a 1D system in the class DIII~\cite{Qi2010PRB}, the $\mathbb{Z}_2$ invariant $\nu_Q \in \{0,1\}$ is given by
\begin{align}\label{Z2}
(-1)^{\nu_Q}
&=\frac{\textrm{Pf}[A(\pi)]}{\textrm{Pf}[A(0)]}\times
\exp \left\{ -\frac{1}{2}\int_{0}^{\pi} d\Phi_z \frac{\partial}{\partial \Phi_z} \log \det[A(\Phi_z)] \right\} \nonumber \\
&=\frac{\mathrm{Pf}[A(\pi)]}{\mathrm{Pf}[A(0)]} \sqrt{\frac{\det[A(0)]}{\det[A(\pi)]}},
\end{align}
where $\textrm{Pf}[C]$ denotes the Pfaffian of an antisymmetric matrix $C$, which is well defined for $A(0)$ and $A(\pi)$.
Since $\det[A]=\textrm{Pf}[A]^2$, $(-1)^{\nu_Q}=\pm 1$ so that $\nu_Q$ is quantized to be $0$ or $1$.
Note that it is required that the phase of $\det[A(\Phi_z)]$ changes continuously as $\Phi_z$ varies from
$0$ to $\pi$.

To numerically evaluate the matrices $A(\Phi_z)$ and the $\mathbb{Z}_2$ invariant, we also need to determine the gauge of the occupied eigenstates.
To fix the gauge, we evaluate the sewing matrix $B(\Phi_z)$ for the time-reversal operator ${T}$ defined as
\begin{equation}
[B(\Phi_z)]_{mn}=\langle \psi_m(-\Phi_z)| {T} |\psi_n(\Phi_z)\rangle,
\end{equation}
where $m$ and $n$ run over the indices of occupied eigenstates. Similarly to $A(\Phi_z)$,
the unitary matrix $B(\Phi_z)$ is also antisymmetric at both $\Phi_z=0$ and $\Phi_z=\pi$.
Numerically,
we enforce the constraint on the eigenstates $\{|\psi_n(\Phi_z=0/\pi)\rangle\}$ such that both $B(0)$ and $B(\pi)$ take the form of
\begin{equation}\label{gaugeR}
B(\Phi_z=0/\pi)=I_{N_{occ}/2}\otimes i\sigma_y,
\end{equation}
where
$i\sigma_y
$ is the sewing matrix between each Kramers pair of the occupied eigenstates at either $\Phi_z=0$ or $\Phi_z=\pi$.
Because the first-order topology of the system is trivial,
there is no obstruction to satisfy the constraint.
In fact, we can also choose
the basis of $\{|\psi_n(\Phi_z)\rangle\}=\{|\psi_1(\Phi_z)\rangle,|\psi_2(\Phi_z)\rangle,\cdots,|\psi_{N_{occ}}(\Phi_z)\rangle\}$
and the basis of $\{|\psi_n(-\Phi_z)\rangle\}=\{T|\psi_2(\Phi_z)\rangle,-T|\psi_1(\Phi_z)\rangle,\cdots,
T|\psi_{N_{occ}}(\Phi_z)\rangle,-T|\psi_{N_{occ}-1}(\Phi_z)\rangle\}$ so that
\begin{equation}
B(\Phi_z)=I_{N_{occ}/2}\otimes i\sigma_y.
\end{equation}
We can also deduce the relation that
\begin{equation} \label{AUB}
A(\Phi_z)=U_{Q}(-\Phi_z)B(\Phi_z),
\end{equation}
based on Eq.~(\ref{A_phiz}), the result
\begin{equation}
T|\psi_n(\Phi_z)\rangle= \sum_{m} B_{mn}(\Phi_z) |\psi_m(-\Phi_z)\rangle,
\end{equation}
and the definition for the matrix
\begin{equation}
[U_{Q}(\Phi_z)]_{mn}=\langle \psi_m(\Phi_z)| \hat{U}_{Q} |\psi_n(\Phi_z)\rangle.
\end{equation}

With the above gauge constraint and Eq.~(\ref{AUB}), we obtain $\det(A(\Phi_z))=\det(U_Q(-\Phi_z))=\det(U_Q(\Phi_z))$, leading to
\begin{eqnarray} \label{SAA}
\sqrt{\frac{\det[A(0)]}{\det[A(\pi)]}}
&=&\exp \left\{ -\frac{1}{2}\int_{0}^{\pi} d\Phi_z \frac{\partial}{\partial \Phi_z} \log \det(U_Q(\Phi_z)) \right\} \\
&=&\exp \left\{ -\frac{i}{2}\int_{\Phi_z=0}^{\Phi_z=\pi} d\theta  \right\} \sqrt{\frac{r(\Phi_z=0)}{r(\Phi_z=\pi)}},
\end{eqnarray}
where we have used $\det(U_Q(\Phi_z))=r(\Phi_z) e^{i\theta(\Phi_z)}$ with $r(\Phi_z)>0$ and the result
\begin{eqnarray}
U_Q(-\Phi_z)=B(\Phi_z)U_Q^T(\Phi_z)B^\dagger(\Phi_z).
\end{eqnarray}
Recall the definition of the quadrupole moment
\begin{equation}\label{QuadNonInt}
Q_{xy}(\Phi_z)=\frac{1}{2\pi}\text{Im}\log\det(U_Q(\Phi_z))=\frac{\theta(\Phi_z)}{2\pi}.
\end{equation}
Based on it, we can further simplify Eq.~(\ref{SAA}) to
\begin{eqnarray}
\sqrt{\frac{\det[A(0)]}{\det[A(\pi)]}}
&=&\exp \left\{ -i\pi\int_{\Phi_z=0}^{\Phi_z=\pi} d Q_{xy}(\Phi_z)  \right\} \sqrt{\frac{r(\Phi_z=0)}{r(\Phi_z=\pi)}} \\
&=& \exp \left\{ -i\pi[Q_{xy}(\Phi_z=\pi)-Q_{xy}(\Phi_z=0)]  \right\} \sqrt{\frac{r(\Phi_z=0)}{r(\Phi_z=\pi)}},
\end{eqnarray}
where $Q_{xy}(\Phi_z=\pi)$ should be continuously connected to $Q_{xy}(\Phi_z=0)$.
While we obtain the result based on a specific gauge, the quadrupole moment $Q_{xy}(\Phi_z)$ is gauge
independent up to an integer (i.e., the difference between values of the quadrupole moments for two different
gauges at a fixed $\Phi_z$ can only be an integer), meaning that $Q_{xy}(\Phi_z)$ can be easily made
continuous with respect to $\Phi_z$. Since the term $\sqrt{\frac{r(\Phi_z=0)}{r(\Phi_z=\pi)}}$ only contributes a positive factor, we
can discard it and simplify the formulation of the $\mathbb{Z}_2$ index to
\begin{align}
(-1)^{\nu_Q}
&=\sgn \left\{ \frac{\textrm{Pf}[A(\pi)]}{\textrm{Pf}[A(0)]}
\exp \left[ -i\pi\int_{0}^{\pi} d\Phi_z \frac{\partial Q_{xy}}{\partial{\Phi_z}} \right] \right\}, \\ \label{Z2IndexFinal}
&=\sgn \left\{ \frac{\textrm{Pf}[A(\pi)]}{\textrm{Pf}[A(0)]}
\exp \left\{ -i\pi[Q_{xy}(\Phi_z=\pi)-Q_{xy}(\Phi_z=0)]  \right\} \right\},
\end{align}
where the matrices $A(0)$ and $A(\pi)$ are obtained by choosing the gauge required in Eq.~(\ref{gaugeR}).

In our case, we find that $Q_{xy}(\Phi_z)=Q_{xy}(\Phi_z=0)$ for all $\Phi_z$ and the $\mathbb{Z}_2$ invariant
only depends on the Pfaffian as shown in Fig.~\ref{figS4}(a).

\begin{figure}[t]
\includegraphics[width=5.5in]{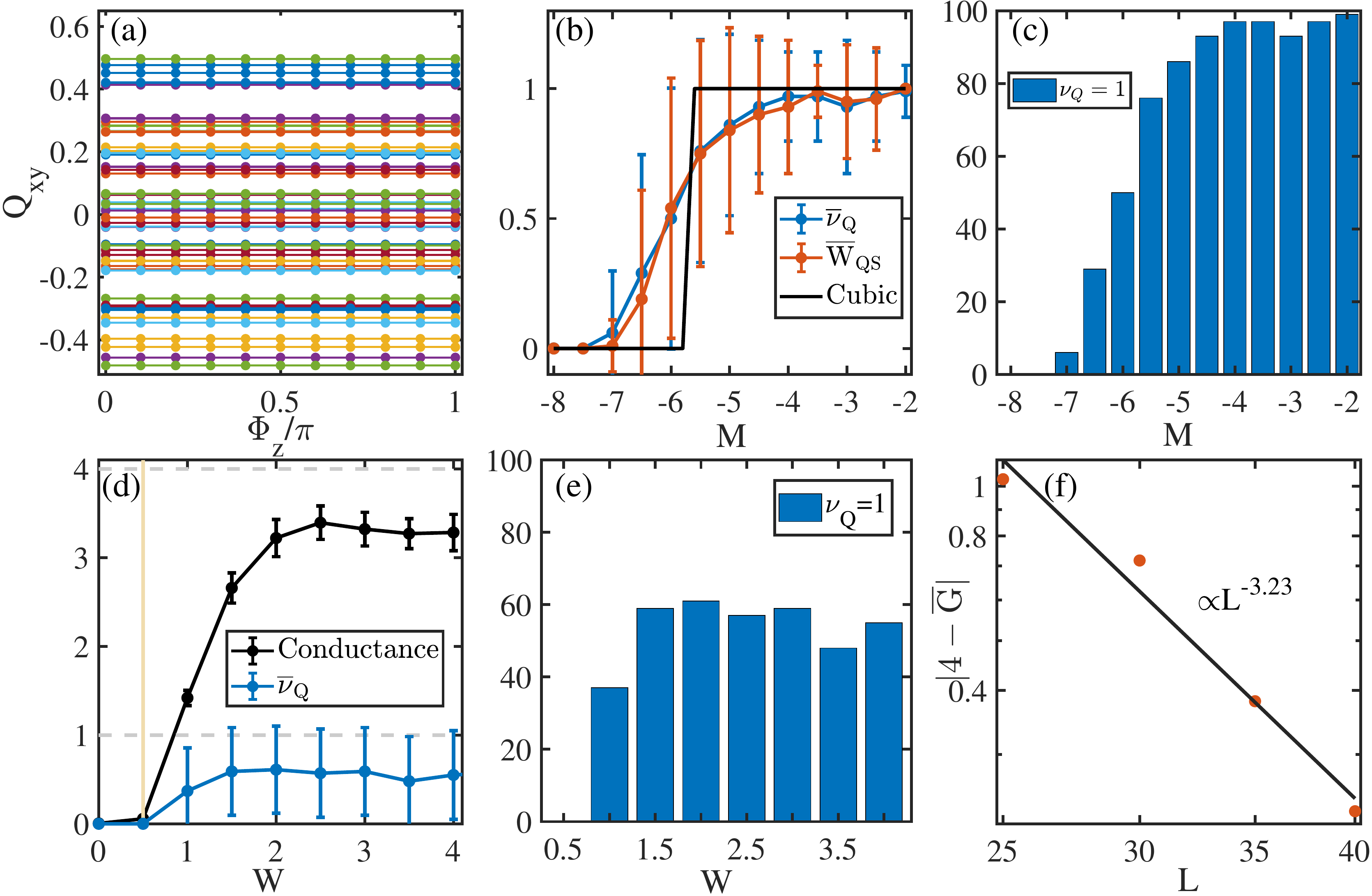}
\caption{(Color online)
(a) Illustration of constant quadrupole moments with respect to $\Phi_z$ when $M=-4$ for 50 disorder realizations.
(b) Configuration averaged $\mathbb{Z}_2$ invariants $\overline{\nu}_Q$ (blue line) and spin quadrupole moment winding numbers
$\overline{W}_{QS}$ (red line) with standard deviations. The black line depicts the $\mathbb{Z}_2$ invariant and the spin
quadrupole moment winding number $W_{QS}$ in a cubic lattice geometry, which gives identical results for both invariants.
(c) Plot of the number of disorder samples with $\nu_{Q}=1$ in 100 disorder realizations with respect to $M$. In (b)-(c), the system parameters are the same as in Fig. (4) in the main text.
(d) Configuration averaged longitudinal conductances (black line) and $\mathbb{Z}_2$ invariants $\overline{\nu}_Q$ (blue line)
with standard deviations versus the structural disorder strength $W$
for Hamiltonian (4) in the main text with $M=-6$. The black and blue lines correspond to a system with size $L=30$ and $L=12$, respectively.
(e) Plot of the number of disorder samples with $\nu_Q=1$ in 100 disorder realizations for the blue line in (d).
(f) $|4-\overline{G}|$ versus system size $L$,
showing $|4-\overline{G}| \propto L^{-3.23}$, indicating that $\overline{G}$ approaches $4 e^2/h$ in the thermodynamic limit. Here, $M=-6$ and
$W=4$.
}
\label{figS4}
\end{figure}

\subsection{B. More numerical results}
In Fig. 4 in the main text, we have plotted the $\mathbb{Z}_2$ invariants for both cubic and amorphous systems,
which are in good agreement with the conductance results. Here, to show the fluctuations, we further give the plot of the invariant
with a standard deviation and the number of configurations with $\nu_Q=1$ in Fig.~\ref{figS4}.
We also remark that the topological phase transition corresponds to a bulk energy gap closing since we here only
consider a system with an average $\hat{C}_4$ symmetry for simplicity. Without the average symmetry,
one can still use the $\mathbb{Z}_2$ invariant to characterize the SOTI with TRS.
In addition, we plot the longitudinal conductance and the $\mathbb{Z}_2$ invariant with respect to the
structural disorder strength $W$, showing their abrupt change from zero to nonzero values; it indicates
the structural disorder driven SOTI with TRS.
Yet, we see strong fluctuations. There are two reasons for their occurrence. One reason is that the system size that
we consider is too small. The other is that the system parameter is close to the critical point so that its
energy gap is very small. To further illustrate the finite-size effects, we plot the configuration averaged conductance as
a function of system sizes, showing $|4-\overline{G}|\propto L^{-3.23}$, which indicates that
$\overline{G}$ should approach the quantized conductance of $4e^2/h$ in the thermodynamic limit.

\section{S-4. A spin quadrupole moment winding number}

In this section, we generalize the quadrupole moment winding number defined in the main text to a spin quadrupole moment winding number for a system with the conservation of a spin (or pseudospin) degree of freedom.

\subsection{A. Without time-reversal symmetry}

For these systems, we can decompose the total Hamiltonian into two subspaces corresponding to different eigenvalues of the conserved spin component,
e.g., the pseudospin denoted by $s_y$ in the Hamiltonian $\hat{H}_h$ with $t_3=0$ in the main text.
Then in each spin eigenspace, we can define the winding number of the quadrupole moment as
\begin{equation}\label{Wq_mb}
W_Q^{(s)}=\int_{0}^{2\pi} d\Phi_z \frac{\partial Q_{xy}^{(s)}(\Phi_z)}{\partial \Phi_z},
\end{equation}
where $Q_{xy}^{(s)}(\Phi_z)$ is the quadrupole moment of occupied states in the subspace with the eigenvalue of $s_y$ being $s$
for the Hamiltonian under the flux $\hat{H}(\Phi_z)$. $Q_{xy}^{(s)}$ is calculated through~\cite{Cho2019PRB,Wheeler2019PRB}
\begin{equation}
Q_{xy}^{(s)}(\Phi_z)
=\frac{1}{2\pi}\mathrm{Im} \log \langle \Psi_G^{(s)}(\Phi_z)| e^{i2\pi \hat{q}_{xy}} |\Psi_G^{(s)}(\Phi_z)\rangle,
\end{equation}
where $\hat{q}_{xy}=\sum_{\bf r} xy \hat{n}({\bf r})/(L_x L_y)$ with $\hat{n}({\bf r})$ being the electron number operator at site ${\bf r}$,
and $|\Psi_G^{(s)}(\Phi_z)\rangle$ is the many-body ground state of the subspace Hamiltonian $\hat{H}^{(s)}(\Phi_z)$ with the
spin (or pseudospin) eigenvalue $s$.

For noninteracting electrons, the many-body ground state $|\Psi_G \rangle$ can be represented as the Slater determinant of occupied single-particle states so that the quadrupole moment can be formulated by Eq.~(\ref{QuadNonInt}). For the quadrupole moment in a spin (or pseudospin) eigenspace
with eigenvalue of $s$,
we can recast it into the form of
\begin{equation}
Q_{xy}^{(s)}(\Phi_z)=\frac{1}{2\pi}\mathrm{Im} \log \det U_Q^{(s)}(\Phi_z),
\end{equation}
where the matrix $U_Q^{(s)}(\Phi_z)$ is defined as
\begin{equation}
[U_Q^{(s)}(\Phi_z)]_{mn}=\langle \psi_m^{(s)}(\Phi_z) |\hat{U}_{Q}|\psi_n^{(s)}(\Phi_z)\rangle
\end{equation}
with $|\psi_n^{(s)}(\Phi_z)\rangle$ denoting the $n$th occupied single-particle eigenstate in the subspace with the spin (or pseudospin)
eigenvalue $s$.

In this case, the system is classified as $\mathbb{Z}\times\mathbb{Z}$ as there are two winding numbers associated with each subspace.

\subsection{B. With time-reversal symmetry}

If the system has the time-reversal symmetry with $T^2=-1$ in addition to a conserved half spin (or pseudospin) degree of freedom,
e.g., $s_y$, which
is antisymmetric with the time-reversal operator,
we can prove that
the spin quadrupole moment winding numbers for two spin (or pseudospin) subspaces related by the time-reversal operator $\hat{T}$
are opposite in sign, i.e., $W_Q^{(s)}=-W_Q^{(-s)}$ with the spin index $s=\pm 1$. In this case, the system is classified as $\mathbb{Z}$.

To be concrete, because of the time-reversal symmetry, we can always choose a gauge such that  $T|\psi_n^{(s)}(\Phi_z)\rangle=|\psi_n^{(-s)}(-\Phi_z)\rangle$, leading to
\begin{equation}
U_Q^{(s)}(\Phi_z)=[U_Q^{(-s)}(-\Phi_z)]^T,
\end{equation}
which is derived through
\begin{eqnarray}
[U_Q^{(s)}(\Phi_z)]_{mn}
&=&\langle \psi_m^{(s)}(\Phi_z) |T^{-1}(\hat{U}_{Q})^*T|\psi_n^{(s)}(\Phi_z)\rangle \\
&=&\langle T\psi_m^{(s)}(\Phi_z) |(\hat{U}_{Q})^*|T\psi_n^{(s)}(\Phi_z)\rangle^* \\
&=& \langle T\psi_n^{(s)}(\Phi_z) |\hat{U}_{Q}|T\psi_m^{(s)}(\Phi_z)\rangle  \\
&=& \langle \psi_n^{(-s)}(-\Phi_z) |\hat{U}_{Q}|\psi_m^{(-s)}(-\Phi_z)\rangle \\
&=& [U_Q^{(-s)}(-\Phi_z)]_{nm}.
\end{eqnarray}
We thus have $\det(U_Q^{(s)}(\Phi_z))=\det(U_Q^{(-s)}(-\Phi_z))$, yielding
\begin{equation}
Q_{xy}^{(s)}(\Phi_z)=Q_{xy}^{(-s)}(-\Phi_z).
\end{equation}
The above relation also guarantees the degeneracy of the quadrupole moments in two subspaces at $\Phi_z=0,\pi$.

We can further prove that
\begin{equation}
W_Q^{(s)}=-W_Q^{(-s)}
\end{equation}
by
\begin{align}
W_Q^{(s)}&=\int_{0}^{2\pi} d\Phi_z \frac{\partial Q_{xy}^{(s)}(\Phi_z)}{\partial \Phi_z} \nonumber \\
&=\int_{0}^{2\pi} d\Phi_z \frac{\partial Q_{xy}^{(-s)}(-\Phi_z)}{\partial \Phi_z} \nonumber \\
&=-\int_{-2\pi}^{0} d\Phi_z \frac{\partial Q_{xy}^{(-s)}(\Phi_z)}{\partial \Phi_z} \nonumber \\
&=-W_Q^{(-s)},
\end{align}
where we have used the property that $Q_{xy}^{(s)}(\Phi_z+2\pi)=Q_{xy}^{(s)}(\Phi_z)$
because ${H}^{(s)}(\Phi_z+2\pi)$ and ${H}^{(s)}(\Phi_z)$ are related by a unitary transformation so that their quadrupole moments are equal.

We now define a spin winding number of the quadrupole moment as
\begin{eqnarray}
W_{QS}&=&\frac{1}{2}\left( W_Q^{(s=1)}-W_Q^{(s=-1)} \right) \label{sWd1} \\
&=&W_Q^{(s=1)}.
\end{eqnarray}

In Figs.~\ref{figS5}(a) and (b), we plot the spin quadrupole moments $Q_{xy}^{(s)}$ with respect to the flux for
the time-reversal symmetric Hamiltonian $\hat{H}_h$ with spin conservation ($t_3=0$) in the main text.
Here, we consider two configurations of amorphous lattices of size $L=20$ in different regimes.
We can see that for $M=-4$, the quadrupole moments in different pseudospin eigenspaces wind in opposite directions and have a nontrivial
spin winding number,
i.e., $W_{QS}=1$, which characterizes the time-reversal symmetric SOTI phase with $s_y$ symmetry.
In contrast, when $M=-8$, $W_Q^{(s=1)}=W_Q^{(s=-1)}=0$, indicating a trivial insulating state.

\begin{figure}[t]
\includegraphics[width=3.5in]{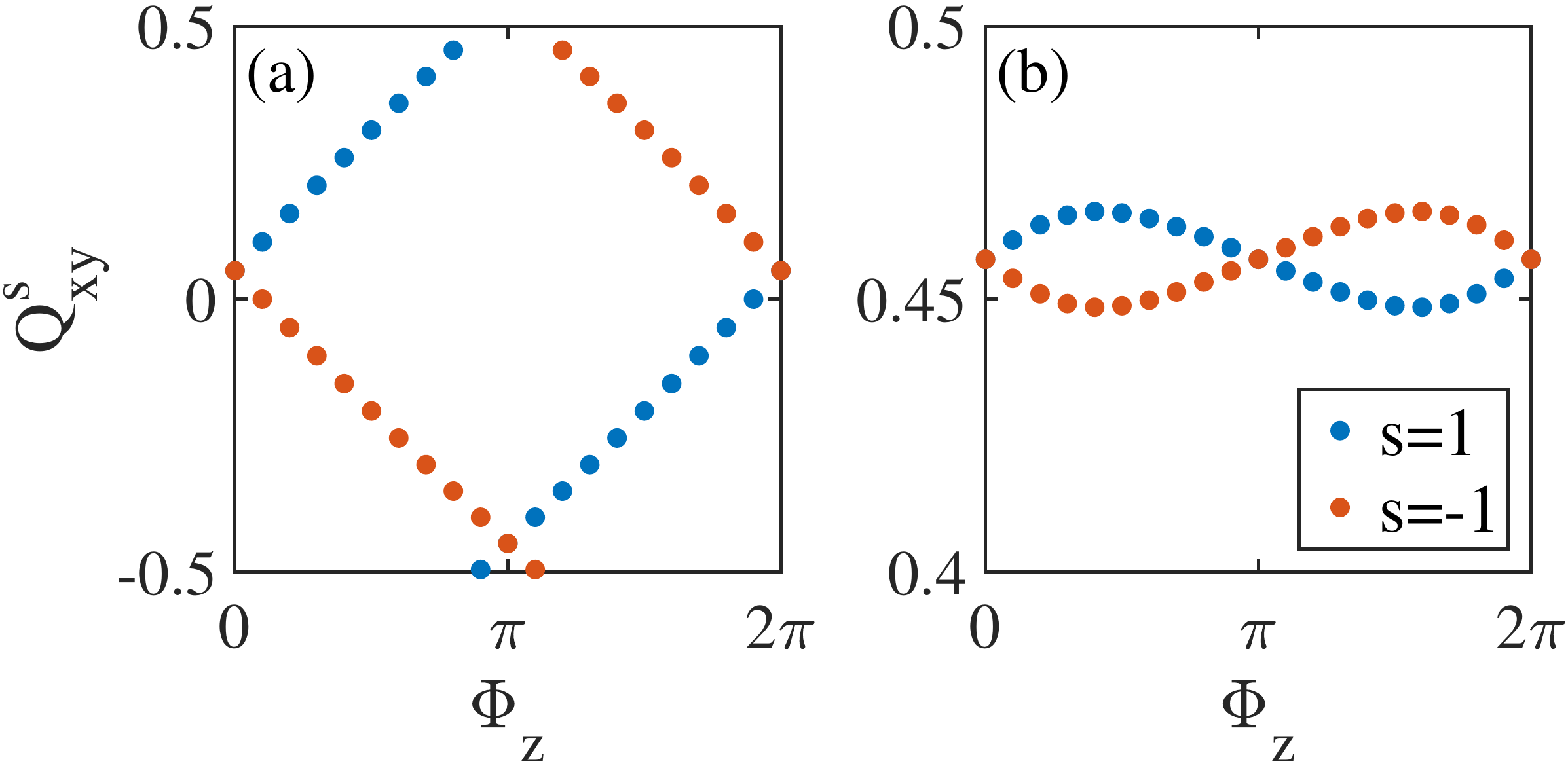}
\caption{(Color online)
The winding of spin quadrupole moments with respect to a flux for the time-reversal symmetric
Hamiltonian $\hat{H}_h$ with spin conservation ($t_3=0$) in amorphous systems.
(a) A nontrivial winding for a typical configuration with $M=-4$ in the topologically nontrivial regime.
(b) A trivial winding for a typical configuration with $M=-8$ in the topologically trivial regime.
}
\label{figS5}
\end{figure}

When the pseudospin symmetry is broken, the classification becomes $\mathbb{Z}_2$, and the $\mathbb{Z}_2$ topological
index can be numerically evaluated based on Eq.~(\ref{Z2IndexFinal}).
In fact, we find that when the pseudospin symmetry is not strongly broken, we can still evaluate the spin
quadrupole moment winding number.

Specifically,
we define a projected spin operator as
\begin{equation}
\hat{P}_s(\Phi_z)=\hat{P}(\Phi_z){s}_y\hat{P}(\Phi_z),
\end{equation}
where $\hat{P}=\sum_{n=1}^{N_{occ}} |\psi_n(\Phi_z)\rangle \langle \psi_n(\Phi_z) |$ is the projection operator to the occupied subspace.
When $t_3=0$, ${s_y}$ is conserved, and thus the nonzero eigenvalues of $\hat{P}_s$ are equal to either $1$ or $-1$.
In the presence of terms breaking the pseudospin symmetry,
if the breaking is not strong, it is possible that the nonzero eigenvalues of $\hat{P}_s(\Phi_z)$
can still be divided into upper and lower bands around $\pm 1$ with respect to $\Phi_z$;
these two bands are separated by a finite gap. Let $\{|\phi_n^{(\pm)}\rangle\}$ be a pair of sets consisting of the corresponding
eigenstates for each band (note that the eigenstates should not contain any contribution from the unoccupied bands).
In this case, we can use
\begin{equation}
[\tilde{U}_Q^{(s)}(\Phi_z)]_{mn}=\langle \phi_m^{(s)}(\Phi_z) |\hat{U}_{Q}|\phi_n^{(s)}(\Phi_z)\rangle
\end{equation}
with $s=\pm$ to calculate the winding number $\tilde{W}_Q^{(s)}$ of the quadrupole moment for each band as well as their spin winding number based on
the following equation
\begin{eqnarray}
\tilde{W}_{QS}&=&\frac{1}{2}\left( \tilde{W}_Q^{(s=1)}-\tilde{W}_Q^{(s=-1)} \right) \mod 2.
\end{eqnarray}

In Fig.~\ref{figS4}(b), we plot the spin winding number for both regular and amorphous lattices for a system with $t_3=1$ breaking
the pseudospin symmetry, showing excellent agreement with the result of the $\mathbb{Z}_2$ topological index and
the longitudinal conductance.

\section{S-5. The relation between the $\mathbb{Z}_2$ invariant and the spin quadrupole moment winding number}

In this section, we discuss the relation between the $\mathbb{Z}_2$ invariant $\nu_Q$ and the spin quadrupole moment winding $W_{QS}$ for
a system with both the TRS and the spin (or pseudospin) conservation,
which is analogous to the relation between a $\mathbb{Z}_2$ invariant and a spin Chern number for a two-dimensional quantum spin Hall insulator~\cite{Fu2006PRB}.

Consider a system with both TRS with ${T}^2=-1$
and a spin (or pseudospin) symmetry, say, $s_y$ in Hamiltonian (4) when $t_3=0$ in the main text.
We now write the matrix $A(\Phi_z)$ (\ref{A_phiz}) in the basis of spin-up (or pseudospin-up) $\{|\psi_n^{(1)}(\Phi_z)\rangle\}$ and
spin-down (or pseudospin-down) $\{|\psi_n^{(-1)}(\Phi_z)\rangle\}$ eigenstates of the conserved spin as
\begin{equation}
A(\Phi_z)=
\begin{pmatrix}
  A^{(1,-1)}(\Phi_z) & A^{(1,-1)}(\Phi_z) \\
  A^{(-1,1)}(\Phi_z) & A^{(-1,-1)}(\Phi_z) \\
\end{pmatrix},
\end{equation}
where
\begin{equation}
[A^{(s_1,s_2)}(\Phi_z)]_{mn}=\langle \psi^{(s_1)}_m(-\Phi_z)| \hat{U}_{Q} {T} |\psi^{(s_2)}_n(\Phi_z)\rangle.
\end{equation}
Since the time-reversal operator ${T}$ transforms spin-up to spin-down (and vise versa),
we can choose the gauge of eigenstates such that
$|\psi^{(\pm)}_n(-\Phi_z)\rangle=\pm {T}|\psi^{(\mp)}_n(\Phi_z)\rangle$.
Using this gauge and considering the fact that $\hat{U}_Q$ only acts on spatial degrees of freedom,
the matrix $A(\Phi_z)$ takes the form of
\begin{equation}
A(\Phi_z)=
\begin{pmatrix}
  0 & U_Q^{(+1)}(-\Phi_z) \\
  -U_Q^{(-1)}(-\Phi_z) & 0 \\
\end{pmatrix},
\end{equation}
where
\begin{align}
[U_Q^{(+1)}(\Phi_z)]_{mn}&=\langle \psi^{(+1)}_m(\Phi_z)| \hat{U}_{Q} |\psi^{(+1)}_n(\Phi_z) \rangle \\
&=\langle \hat{T}\psi^{(-1)}_m(-\Phi_z) | \hat{U}_{Q} |\hat{T} \psi^{(-1)}_n(-\Phi_z) \rangle \\
&= \langle \psi^{(-1)}_m(-\Phi_z) | \hat{U}_{Q} | \psi^{(-1)}_n(-\Phi_z) \rangle^{*} \\
&=\langle \psi^{(-1)}_n(-\Phi_z) | \hat{U}_{Q} | \psi^{(-1)}_m(-\Phi_z) \rangle \\
&=[U_Q^{(-1)}(-\Phi_z)]_{nm}
\end{align}
so that $U_Q^{(+1)}(\Phi_z) = [U_Q^{(-1)}(-\Phi_z)]^T$ leading to the antisymmetric matrix $A(\Phi_z)$ at $\Phi_z=0,\pi$.

We now evaluate the $\mathbb{Z}_2$ topological invariant (\ref{Z2})
\begin{align}
(-1)^{\nu_Q}&= \frac{\mathrm{Pf}[A(\pi)]}{\mathrm{Pf}[A(0)]}\times
\exp \left\{ -\frac{1}{2}\int_{0}^{\pi} d\Phi_z \frac{\partial}{\partial \Phi_z} \log \det[A(\Phi_z)] \right\} \nonumber \\
&= \frac{\det U_Q^{(+1)}(\pi)}{\det U_Q^{(+1)}(0)} \times \exp \left\{ -\frac{1}{2}\int_{0}^{\pi} d\Phi_z \frac{\partial}{\partial \Phi_z}
\left(\log \det[U_Q^{(+1)}(-\Phi_z)] + \log \det[U_Q^{(-1)}(-\Phi_z)] \right) \right\} \nonumber \\
&=\exp \left\{ \frac{1}{2}\int_{0}^{\pi} d\Phi_z \frac{\partial}{\partial \Phi_z}
\left(2\log \det[U_Q^{(+1)}(\Phi_z)] - \log \det[U_Q^{(+1)}(-\Phi_z)] - \log \det[U_Q^{(-1)}(-\Phi_z)] \right) \right\} \nonumber \\
&=\exp \left\{ \frac{1}{2}\int_{0}^{\pi} d\Phi_z \frac{\partial}{\partial \Phi_z}
\left(\log \det[U_Q^{(+1)}(\Phi_z)] - \log \det[U_Q^{(+1)}(-\Phi_z)] \right) \right\} \nonumber \\
&=\exp \left\{ \frac{1}{2}\int_{-\pi}^{\pi} d\Phi_z \frac{\partial}{\partial \Phi_z}
\log \det[U_Q^{(+1)}(\Phi_z)] \right\}
= \exp \left\{ - \frac{1}{2}\int_{-\pi}^{\pi} d\Phi_z \frac{\partial}{\partial \Phi_z}
\log \det[U_Q^{(-1)}(\Phi_z)] \right\} \nonumber \\
&=\exp\left\{ i\pi W_Q^{(+)} \right\}=\exp\left\{-i\pi W_Q^{(-)}\right\} \nonumber \\
&=\exp\left\{i\pi \frac{1}{2} \left( W_Q^{(+)}-W_Q^{(-)} \right) \right\} \nonumber \\
&=\exp\left( i\pi W_{QS} \right),
\end{align}
where we have used the property that for an antisymmetric matrix
$A=
\begin{pmatrix}
  0 & M \\
  -M^{T} & 0 \\
\end{pmatrix}
$, $\mathrm{Pf}(A)= (-1)^{n(n-1)/2} \det(M)$ with $n$ being the dimension of the matrix $M$.
We hence conclude that the $\mathbb{Z}_2$ topological index $\nu_Q$ gives the parity of the spin quadrupole moment
winding number $W_{QS}$ for a system with both TRS and a spin (or pseudospin) symmetry.
\end{widetext}

\end{document}